\documentclass[longauth]{aa}

\usepackage{graphicx}
\usepackage{gensymb}
\usepackage{txfonts}
\usepackage{subcaption}

\def\baysea{\texttt{BaySeAGal}}

\newcommand{\pycasso}{{\sc p}y{\sc casso}}          	 
\newcommand{\starlight}{{\sc starlight}}

\usepackage{xcolor}

\newcommand{\jplus}{J-PLUS}

\newcommand{\jp}{J-PAS}
\newcommand{\js}{J-spectra}
\newcommand{\mjp}{miniJPAS}

\def\sext{\texttt{SExtractor}}

\def\Py{\texttt{Python}}
\def\PyDJ{\texttt{Py2DJPAS}}

\def\magauto{\texttt{MAG\_AUTO}}

\def\magpetro{\texttt{MAG\_PETRO}}

\def\kron{\texttt{KRON\_RADIUS}}
\def\petror{\texttt{PETRO\_RADIUS}}
\def\infot{\texttt{info\_table}}

\def\rb{$r_\mathrm{SDSS}$}
\def\gb{$g_\mathrm{SDSS}$}
\def\ib{$i_\mathrm{SDSS}$}

\def\ujp{$u_\mathrm{JPAS}$}

\begin{document}

   \title{The miniJPAS survey:}

   \subtitle{Exploring the spatially-resolved capabilities of the J-PAS survey with \PyDJ }

   \author{J.E. Rodríguez-Martín.
          \inst{\ref{IAA}}
          \and
          L.A. Díaz-García,
          \inst{\ref{IAA}}
          \and
          R.M. González Delgado, 
          \inst{\ref{IAA}}     
          \and
          G. Martínez-Solaeche,
          \inst{\ref{IAA}}
          \and
          R. García-Benito,
          \inst{\ref{IAA}}
          A. de Amorim,
          \inst{\ref{UFSC}}
          \and
          J. Thainá-Batista
           \inst{\ref{UFSC}}
          \and
          R. Cid Fernandes
           \inst{\ref{UFSC}}
           \and 
           I. Márquez
           \inst{\ref{IAA}}
           \and
            A. Fernández-Soto
            \inst{\ref{UC}, \ref{UV}}
            \and
            I. Breda
            \inst{\ref{UniVie}}
           \and
           R. Abramo
           \inst{\ref{USP-IF}}
           \and
           J. Alcaniz
           \inst{\ref{ON}}
           \and
           N. Benítez
           \and
           S. Bonoli
           \inst{\ref{DIPC}, \ref{IKERBASQUE}}
           \and
           S. Carneiro
           \inst{\ref{ON}}
           \and
           A. J. Cenarro \inst{\ref{CEFCA},\ref{CEFCACSIC}}
           \and
           D. Cristóbal-Hornillos \inst{\ref{CEFCA}}
           \and
           R. A. Dupke \inst{\ref{ON}}
           \and
           A. Ederoclite \inst{\ref{CEFCA},\ref{CEFCACSIC}}
           \and
           A. Hernán-Caballero \inst{\ref{CEFCA},\ref{CEFCACSIC}}
           \and
           C. Hernández-Monteagudo \inst{\ref{IAC}, \ref{ULL}}
           \and
           C. López-Sanjuan \inst{\ref{CEFCA},\ref{CEFCACSIC}}
           \and
           A. Marín-Franch \inst{\ref{CEFCA},\ref{CEFCACSIC}}
           \and
           C. Mendes de Oliveira \inst{\ref{USP}}
           \and
           M. Moles \inst{\ref{CEFCA}}
           \and
           L. Sodré \inst{\ref{USP}}
           \and
           K. Taylor \inst{\ref{instruments4}}
           \and
           J. Varela \inst{\ref{CEFCA}}
           \and
           H. Vázquez Ramió \inst{\ref{CEFCA},\ref{CEFCACSIC}}
          }

   \institute{Instituto de Astrofísica de Andalucía (IAA-CSIC), P.O.~Box 3004,
                18080 Granada, Spain\\
              \email{julioeroma@iaa.csic.es, julioeroma@gmail.com }
              \label{IAA}
         \and
             Departamento de F\'{\i}sica, Universidade Federal de Santa Catarina, P.O.~Box 476, 88040-900, Florian\'opolis, SC, Brazil \label{UFSC} 
        \and 
            Instituto de Física de Cantabria (CSIC-UC), Avda. Los Castros s/n, 39005 Santander, Spain \label{UC}
        \and
            Unidad Asociada “Grupo de Astrofísica Extragaláctica y Cosmología”, IFCA-CSIC/Universitat de València, València, Spain \label{UV}
        \and
            Dep. of Astrophysics, University of Vienna, Türkenschanzstraße 17, 1180, Vienna, Austria \label{UniVie}        
        \and
            Instituto de F\'isica, Universidade de S\~ao Paulo, Rua do Mat\~ao 1371, CEP 05508-090, S\~ao Paulo, Brazil\label{USP-IF}
        \and
            Observatório Nacional – MCTI (ON), Rua Gal. José Cristino 77, São Cristóvão, 20921-400 Rio de Janeiro, Brazil \label{ON}
        \and
            Donostia International Physics Center (DIPC), Manuel Lardizabal Ibilbidea 4, San Sebastián, Spain \label{DIPC}
        \and
            IKERBASQUE, Basque Foundation for Science, 48013 Bilbao, Spain \label{IKERBASQUE}
        \and
            Instituto de Física, Universidade Federal da Bahia, 40210-340 Salvador, BA, Brazil \label{IF-UFB}
        \and
            Centro de Estudios de F\'isica del Cosmos de Arag\'on (CEFCA), Plaza San Juan 1, 44001 Teruel, Spain \label{CEFCA} 
        \and
            Unidad Asociada CEFCA-IAA, CEFCA, Unidad Asociada al CSIC por el IAA, Plaza San Juan 1, 44001 Teruel, Spain. \label{CEFCACSIC}          
        \and
            Instituto de Astrofísica de Canarias (IAC), C/ Vía Láctea, S/N, E-38205, San Cristóbal de La Laguna, Tenerife, Spain \label{IAC}
        \and
            Departamento de Astrofísica, Universidad de La Laguna (ULL), Avenida Francisco Sánchez, E-38206, San Cristóbal de La Laguna, Tenerife, Spain \label{ULL}
        \and
        Universidade de São Paulo, Instituto de Astronomia, Geofísica e Ciências Atmosféricas, Rua do Matão 1226, 05508-090 São Paulo, Brazil \label{USP}
        \and
        Instruments4, 4121 Pembury Place, La Canada Flintridge, CA 91011, USA \label{instruments4}
             }

   \date{Received \today; accepted ?? ??, ??}

  \abstract
{This work presents \PyDJ, a tool developed in \Py\ to automate the analysis of the properties of spatially resolved galaxies in the \mjp \ survey, a 1~deg$^2$ survey that acts as precursor of the Javalambre Physics of the Acclerating Universe Survey (\jp), using the same filter system and telescope, and the J-PAS Pathfinder camera. Our goal is to provide a single code to download the scientific images and tables required for analysis, perform point spread function (PSF) homogenization, automate masking, define apertures, and run an external fitting code for spectral energy distribution (SED) analysis, as well as estimate the equivalent widths of the main optical emission lines via an artificial neural network. 

We select a sample of spatially resolved galaxies in \mjp\ and calculate their magnitudes in all bands to demonstrate that we retrieve the same values as those provided in the \mjp\ catalog using \sext\ with a precision of approximately $10\%$. We show that these measurements are significantly improved by the local estimation of the background for dimmer galaxies and apertures. The PSF homogenization enhances multi-band photometry in the innermost apertures, ensuring consistent apertures across filters that permit generating pseudo-spectra without undesired artifacts or variations due to mismatched photometry. By performing a SED-fitting of the multi-band photometry (\js) within annular apertures over the PSF-homogenized images, we find that the residuals of the fitting remain below $\sim 10\%$, with no significant wavelength-dependent bias for apertures with $S/N > 5$. Thus, our method provides robust photometric measurements, marking the first step toward our goal of studying the spatially resolved properties of galaxies. 
Furthermore, we demonstrate the IFU-like capabilities of \jp\ by analysing the spatially resolved properties of the galaxy 2470-10239 at \( z = 0.078 \) and comparing them with MaNGA data  up to 1 half light radius (HLR), which is the maximum extent for this galaxy in the MaNGA data-cube. We find very good agreement between the photometric and spectroscopic measurements, as well as identical radial profiles of the stellar mass surface density up to 1 HLR. However, our analysis extends further to 4 HLR, where the \mjp\ data have $S/N \sim 5$. This provides evidence of the capability of J-PAS to extend the IFU-like analysis to the outskirts of the galaxy, enabling the study of the processes that drive their evolution at larger galactocentric distances.
}


   \maketitle

\section{Introduction}
Much of our current knowledge about galaxies has been made possible thanks to data from galaxy surveys. The observations carried out by these surveys have provided information on a large number of galaxies at different redshifts. This is key to studying the properties of galaxies at different epochs with a great statistical significance, and to understanding the evolution of galaxies. Some surveys have provided spatially-resolved data for galaxies with larger apparent sizes, which has shed light into the structure of galaxies and the processes that drive their evolution at local scale.

Spectroscopic and photometric surveys are the two primary approaches to observing galaxies, each with distinct advantages and limitations. Spectroscopic surveys, such as the Sloan Digital Sky Survey \citep[SDSS,][]{York2000}, the VIMOS VLT Deep
Survey \citep[VVDS,][]{VVDS2005}, or the VIMOS Public
Extragalactic Redshift Survey \citep[VIPERS,][]{Haines2017}, provide high spectral resolution, enabling accurate spectral fitting and analysis of stellar and ionised gas content \citep[e.g.,][]{Thomas2005,cid2005,SanchezBlazquez2006b}. However, classical spectroscopic surveys can suffer from aperture bias when studying larger galaxies due to limited fibre or slit coverage. In contrast, photometric surveys, such as the Galaxy Evolution from Morphology and SEDS survey \citep[GEMS,][]{Rix2004}, the Dark Energy Survey \citep[DES,][]{Wester2005}, or the Panoramic Survey Telescope and Rapid Response System 1 survey \citep[Pan-STARRS1,][]{Chambers2016},  use filters to map spectral energy distributions (SEDs) with lower spectral resolution but offer advantages such as deeper observations and the absence of aperture and selection biases. Photometric data have been widely used to study stellar population properties in galaxies \citep[see e.g.,][and references therein]{Walcher2011,Luis2015,Luis2019,Luis2019b,Luis2019c,Rosa2021}.

To overcome the limitations of spectroscopic and photometric surveys, particularly the aperture effects in traditional spectroscopy, integral field units (IFUs) were developed. By combining multiple fibres pointed at the same object, IFUs enable the acquisition of spatially resolved data. Notable integral field spectroscopy (IFS) surveys include the Spectroscopic Areal Unit for Research on Optical Nebulae \citep[SAURON,][]{SAURON2001},  ATLAS-3D \citep{Cappellari2011},  the Calar Alto Legacy Integral Field Area survey \citep[CALIFA, ][]{CALIFA2012}, and the Mapping Nearby Galaxies at Apache Point Observatory survey \citep[MaNGA,][]{MANGA2015}. An alternative approach is multiband photometric surveys, which enhance spectral resolution by using a larger number of filters, especially narrowband ones. Examples include the Area Medium Band Redshift Astronomical survey \citep[ALHAMBRA,][]{Moles2008,Molino2014}, the Javalambre Photometric Local Universe Survey \citep[J-PLUS,][]{Cenarro2019}, or the Southern Photometric Local Universe Survey \citep[S-PLUS,][]{SPLUS2019}.

The Javalambre Physics of the Accelerated Universe Astrophysical Survey \citep[J-PAS,][]{Benitez2014} is an ongoing survey that will cover thousands of square degrees in the Northern Hemisphere from the Observatorio Astrofísico de Javalambre \citep[OAJ,][]{Cenarro2014} using the $2.5$~m JST/T250 telescope. One of the most powerful features of \jp\ is its photometric filter system, composed of 56 narrow and medium-band filters, covering the entire optical wavelength range and providing a spectral resolution comparable to very low-resolution spectroscopy. In addition, its effective field of view (FoV) is $4.2$~$\mathrm{deg}^2$, covered by the Javalambre Panoramic Camera \citep[JPCam,][]{Taylor2014,MarinFranch2017}, a 1.2 Gpixel camera with a pixel scale of $0.23$~arcsec~$\mathrm{pix}^{-1}$, utilizing 14 CCDs simultaneously.

The combination of these characteristics makes \jp\ an excellent survey for studying the evolution and properties of galaxies. This has been demonstrated using data from the \mjp\ survey \citep{Bonoli2020}, a 1~deg$^2$ survey that employs the same photometric filter system as \jp. In particular, we have shown the capability of its filter system to fit the spectral energy distribution (SED) of galaxies, allowing us to retrieve stellar population properties with high accuracy and study their evolution over cosmic time \citep[see][]{Rosa2021, Luis2024}. Furthermore, it is also useful for the study of the main emission lines in the covered wavelength range \citep{Gines2021, Gines2022, Breda2024}, and it has been shown to effectively detect quasars through machine learning techniques \citep{Queiroz2023, Rodrigues2023, Gines2023}. Additionally, its large FoV enables the detection of galaxies in groups and clusters without selection bias, as well as for the study of larger galaxies without any preselection bias. This capability has been leveraged to produce galaxy cluster and group catalogues \citep{Doubrawa2023, Maturi2023}, which have been used to investigate the role of the environment in galaxy evolution \citep{Rosa2022, Julio2022}.

Thanks to the combination of its filter system, the imaging of contiguous, large areas of the sky, and its pixel scale, \jp \ will provide 3D-data that are ideal for IFU-like studies. The capability of multiband surveys to perform IFU-like studies have also been shown, particularly using data from ALHAMBRA and J-PLUS surveys, which are closely related to \jp. These spatially-resolved studies include the study of the stellar population properties \citep{SanRoman2018,SanRoman2019}, and the star formation rate of the regions of galaxies \citep{Logrono2019} or the effect of AGNs on the local SFR \citep{Acharya2024}. Furthermore, the large FoV of these surveys allows for studying very large galaxies that would otherwise be outside the limits of the detector (see \citealt{Rahna2025} and \citealt{Julia2025}).

Our ultimate goal in this series of papers is to study the stellar population and emission line properties of spatially resolved galaxies in the \mjp\ survey, in order to demonstrate the IFU-like capabilities of J-PAS. In this first paper, we present and test our tool that automates the analysis of spatially resolved galaxies in the \mjp\ survey, similar to the role played by \pycasso\ \citep{Pycasso2017} for CALIFA or pyPipe3D \citep{Lacerda2022} for MaNGA. In an upcoming paper, we will focus on the results of the SED-fitting and ANN-based estimation for the apertures of the sample of spatially resolved galaxies in groups detected in \mjp.

This paper is structured as follows: in Sect.~\ref{sec:data} we describe the nature of the \mjp \ data, as well as the criteria used to select our sample, in Sect.~\ref{sec:method} we describe the working flow of our code, in Sect.~\ref{sec:results} we compare the photometry obtained with the photometry given in the \mjp \ catalogues, in Sect~\ref{sec:discuss} we discuss our results. All the magnitudes used in this paper are given in the AB magnitude system \cite{Oke1983}. Throughout this paper, we assume a Lambda cold dark matter ($\Lambda$CDM) cosmology with $h = 0.674$, $\Omega _{\mathrm{M}} = 0.315$, $\Omega _\Lambda = 0.685$, based on the latest results by the \cite{Planck2020}.

\section{Data} \label{sec:data}

In this section, we first describe the key details of the \mjp \ survey and its data, as well as the selection criteria for the sample of spatially-resolved galaxies used to illustrate and test our code.

\subsection{The \mjp \ survey}
The \mjp \ survey \citep{Bonoli2020} is a $1$~$\mathrm{deg}^2$ survey that was carried out at the OAJ, using the $2.5$~m  JST/T250 telescope, the same one as J-PAS. The main goals of this survey include testing and showing the potential of the \jp \ photometric filter system, exploring the capabilities of the \jp \ survey, and the first scientific exploitation of the data using this photometric system. 

This system consists of  54 narrow band filters covering the whole 3780--9100~\AA \ optical wavelength range, with a full width at half maximum (FWHM) of $\sim 145$~\AA, equally spaced every 100~\AA. Additionally, there are two intermediate band filters, one covering the UV edge, ($u_\mathrm{JAVA}$, with a central wavelength of 3497~\AA \ and a FWHM of 495~\AA), and another one covering the redder edge, (J1007, with a central wavelength of 9316~\AA \ and a FWHM of 620~\AA). The achieved spectral resolution is $R\sim 60$, equivalent to that of very low resolution spectroscopy. In addition,  four broadband SDSS-like filters were used: \rb, \gb, \ib, and \ujp.

The camera used was the J-PAS Pathfinder  (JPAS-PF) camera.  Unlike the JPCAM, this camera is equipped with a single CCD and with a low-noise $9.2k \times 9.2k$ pixel CCD detector. The resulting FoV is 0.27~$\mathrm{deg}^2$ with a pixel scale of $0.23$~arcsec~$\mathrm{pixel}^{-1}$.

The \mjp \ consists of four pointings across the AEGIS field that partly overlap, resulting in four composite images: miniJPAS-AEGIS1 $(\alpha, \delta) = (214 \degree.2825, 52\degree.5143)$, miniJPAS-AEGIS2 $(\alpha, \delta) = (214\degree.8285, 52\degree.8487)$, miniJPAS-AEGIS3 $(\alpha, \delta) = (215\degree.3879, 53\degree.1832)$, and miniJPAS-AEGIS4 $(\alpha, \delta) = (213\degree.7417, 52\degree.1770)$. These tiles overlap with several surveys, including ALHAMBRA, SDSS, the OSIRIS Tunable Emission Line Object survey \citep[OTELO][]{OTELO2019}, the Hyper Suprime-Cam Subaru Strategic Program \citep[HSC-SSP, ][]{Aihara2018Presentation}, the Extended Goth Strip (EGS), and the Canada–France–Hawaii Telescope Legacy Survey (CFHTLS) Each tile was observed with a minimum of four exposures per filter, with eight exposures per filter for the reddest ones. The exposure times were of 120~s for the narrow band filters and the  $u_\mathrm{JPAS}$ filter, while it was of 30~s for the broad band ones, in order to avoid saturation. These times were defined in order to reach the optimal photometric-redshift depth for the wide range of galaxies at different redshifts, according to the baseline strategy \citep{Benitez2014}.

\subsection{Sample selection} \label{sec:sample}
In order to guarantee that the selected galaxies are suitable for spatially-resolved studies, we impose the following criteria. First, the apparent size of the galaxies in the sample must be larger than the  point spread function (PSF), so we can divide them into independent apertures. We decide to avoid very edge--on galaxies, since the light extracted from them would be a mixture of the light emitted in inner and outer apertures. We use the ellipticity of the aperture as a proxy of their inclination. Lastly, we also desire to avoid artifacts in the images and galaxies whose photometry might be biased by other bright nearby sources. As a consequence, we also need the flags of the catalogues into account. 

Taking all these  considerations into account, our selection criteria can be summarised in the next requirements: (i) the effective radius (R\_EFF) of the galaxy must be larger than $2''$, in order to be able to define at least two apertures, given the typical FWHM of the PSF of the images in \mjp \ \citep[usually $\sim 1-1.25$~arcsec, with all filters below 2~\arcsec, see Fig.~5 in ][for further details]{Bonoli2020}, (ii) the radius obtained as the square root of the isophotal area divided by $\pi$ must be at least two times larger than the FWHM of the PSF, (iii) the ellipticity must be smaller than $0.6$, (iv) the \texttt{MASK\_FLAGS} parameter provided by \sext \ must be 0 for all the filters, in order to avoid galaxies that are outside the window frame, that are near a bright star, or that are masked due to nearby artifact, (v) the $\texttt{FLAGS}$ parameter must not contain the flag 1, which indicates that the object has neighbours, bright and close enough to significantly bias the photometry, or bad pixels  (vi) the $\texttt{CLASS\_STAR}$ parameter must be lower than $0.1$, to filter the maximum number of stellar objects. These criteria gives us a total of 51 galaxies.

\section{Methodology} \label{sec:method}

\subsection{Pre-processing}
The data reduction process of \mjp \ follows the same procedure as that used for J-PLUS \citep{Cenarro2019}. For the single frames, this includes the bias, prescan, and overscan subtraction, the trimming, and the flat field, illumination, and fringing corrections. However, three main issues required special attention during the \mjp \ data reduction process. The first issue was vignetting, caused by the larger size of the CCD compared to the filters, resulting in a strong efficiency gradient. To solve this, regions with low efficiency in the images were trimmed, from $9216 \times 9232$~pixels to $7777 \times 8473$~pixels. The second issue were background patterns: images from \mjp \ exhibited background patterns with strong gradients and variations on timescales of a few minutes. There are two types of patterns. The first type consists of straight patterns, which are caused by the optics of the camera and were corrected via illumination correction. The second type shows circular patterns, likely due to small variations in the central wavelengths of the filters. These ones require a more careful subtraction that is detailed in Appendix B in \cite{Bonoli2020}. The working hypothesis is that these patterns remain the same in images taken close in time, as well as being independent of the sky position. Therefore, a median combination of the images with objects in different positions should preserve the background structure while removing the sources. The last issue was fringing in filters redder than J0470 ($\lambda > 4700$~\AA). To remove these effects, master fringing images were constructed using all available images for each band, since this pattern is very stable across nights. However, some residual pattern may still be present in the final images of a few filters, due to the low number of available images.

 After all this processing, the images were combined using the Astromatic software \texttt{Swarp} \citep{Bertin2010Swarp}. All images were resampled to the pixel size of the camera ($0.23$~arcsec~$\mathrm{pixel}^{-1}$). The images were homogenised using \texttt{PSFEx} \citep{Bertin2011}, and the PSF measurements produced for each image are available at a given position within the image. Lastly, \sext \ was run over the images in order to obtain the different catalogues publicly available.

\subsection{\PyDJ}
\PyDJ \ is our tool based on \Py \ that we have developed with the purpose of automate the steps required for analysing of the properties of the spatially-resolved galaxies in \mjp.  The main steps of the pipeline can be divided in the following blocks: (i) downloading the scientific images and tables, (ii) flux conversion, (iii) image masking, (iv) PSF homogenisation, (v) segmentation and binning of the galaxy,  and (vi) the SED-fitting and line emission analysis. This code will also be useful for analysing the spatially resolved galaxies from \jp \ after adapting these steps to the peculiarities of the survey, such as the more complex PSF models. 

The complete process of the code is detailed in Appendix~\ref{app:code}, along with the libraries and parameter values used for this work. However, we include in this section the most relevant points for the work shown in this paper. The first consideration is that the calculation of the photometry in the apertures used in this work includes a local background correction, as well as a correction estimating the masked pixels. Secondly, throughout this work we use both elliptical and circular apertures and annuli. The elliptical apertures are defined in two ways: first, to match the \magpetro \ and \magauto \ apertures (see \ref{sec:code:photometry} for further details), and second, using steps that increase in the value of the FWHM of the worst PSF. The circular apertures are defined to match the apertures  of $0.8'', 1.0'', 1.2'', 1.5'', 2.0'', 3.0'', 4.0'',$ and $ 6.0''$.

\subsection{SED-fitting and ANN estimation} \label{sec:fit_ANN}
The measurements of the photometry of the apertures obtained with the previous steps are suitable for use as an input for any SED-fitting code. However, in our working version of the code, we have implemented a wrapper that automatically executes \baysea \ \citep{Rosa2021}, a Bayesian SED-fitting code based on a Markov chain Monte Carlo (MCMC)  approach to explore the parameter space and find the set of parameters that  best fit the observed magnitudes. We have used this code in other works, such as \cite{Rosa2022} and \cite{Julio2022}, to obtain the stellar population properties and SFHs of the galaxies in \mjp, and we use the same models and approach as in those works. In this work, we  use the SED-fitting only in the discussion (Sect.~\ref{sec:discuss}) in order to evaluate the residuals of the fitting and the precision that can be achieved with our methodology, leaving the analysis of the stellar population properties for a upcoming paper. In addition, the measurements of the photometry obtained with \PyDJ \ and the results from the SED-fitting are used as an input for the ANN trained by \cite{Gines2021} to estimate the equivalent widths (EW) of the H$\alpha$, H$\beta$, [NII], and [OIII] emission lines. The training set contains  a large, diverse set of spaxels from CALIFA and MaNGA galaxies, ensuring reliable predictions in varied conditions.

\section{Results}\label{sec:results}

In this section, we illustrate the processes of the methodology and compare our magnitude  measurements with those obtained using \sext, which are provided in the \mjp \ catalogues. Our aim is to demonstrate that we obtain reliable measurements of the magnitudes with our methodology, which are necessary for obtaining results regarding the properties of the galaxies. In order to illustrate the process, we focus on the galaxy 2470-10239 ($\alpha =$ $14$~h $15$~min $20.37$~s; $\delta = 52\degree$ $20'$ $45.19''$), which is the only galaxy common to both the \mjp \  sample and the MaNGA survey. This is a bright ($r_{\mathrm{SDSS}} = 14.62$), massive ($M_\star = 10^{11.51}$~$\mathrm{M_\odot}$ from our analysis here), red, elliptical galaxy, at $z=0.074$. It is also the largest galaxy in terms of angular 
in \mjp. This way, we can provide the best possible spatially-resolved analysis until the arrival of \jp \ data for galaxies in the Local Volume. 

\subsection{PSF homogenisation and masking}
\begin{figure*}
    \centering
    \includegraphics[width=\textwidth]{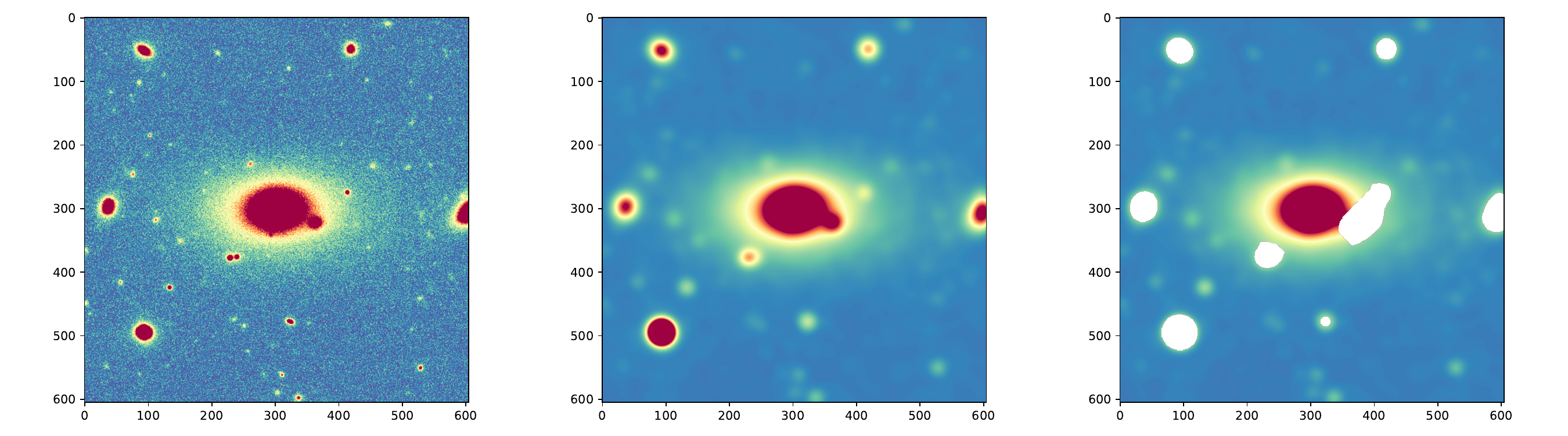}
    \caption[Summary of the image treatment process for the galaxy $2470-10239$: original image, PSF-homogenised image and  masked PSF-homogenised image.]{Summary of the image treatment process. Left panel shows the \rb{} image of the galaxy $2470-10239$ with no treatment. The middle panel shows the \rb{} image after PSF homogenisation. Right panel shows the final image, after PSF homogenisation and applying the sources mask. The colour scale indicates the flux in ADUs.}
    \label{fig:MANGA_psf_mask}
\end{figure*}

Figure~\ref{fig:MANGA_psf_mask} illustrates the entire image processing workflow for the \rb{} image. The effect of the homogenisation on the image is quite clear. The image becomes blurrier, the background and shapes of the sources smooth out, and the smallest sources appear extended. Indeed, the effect of the PSF in the morphology has been extensively studied \citep[see e.g.,][]{PaulinHenriksson2008,Lewis2009,Voigt2010}, but we are primarily concerned with its impact on the photometry. The potential effect on morphology will be considered after recalculating the ellipse parameters. 

The mask used for source detection depends on two parameters: the minimum number of pixels (npix) and the threshold level for detection. After testing various values (see Fig.~\ref{fig:sourcesparam1} for some examples), it was found that setting the threshold too low leads to false detections, while a higher pixel count does not resolve the issue and can associate sources incorrectly. A high threshold might miss dimmer sources. The best compromise found was $rms = 5, \ npix = 30$ using the \rb{} image as the reference, which provided a solid mask for both nearby and smaller sources in the specific case tested. These values were used consistently throughout the work.

\subsection{PSF homogenisation effect on the \js}

\begin{figure*}
    \centering
    \includegraphics[width=0.95\textwidth]{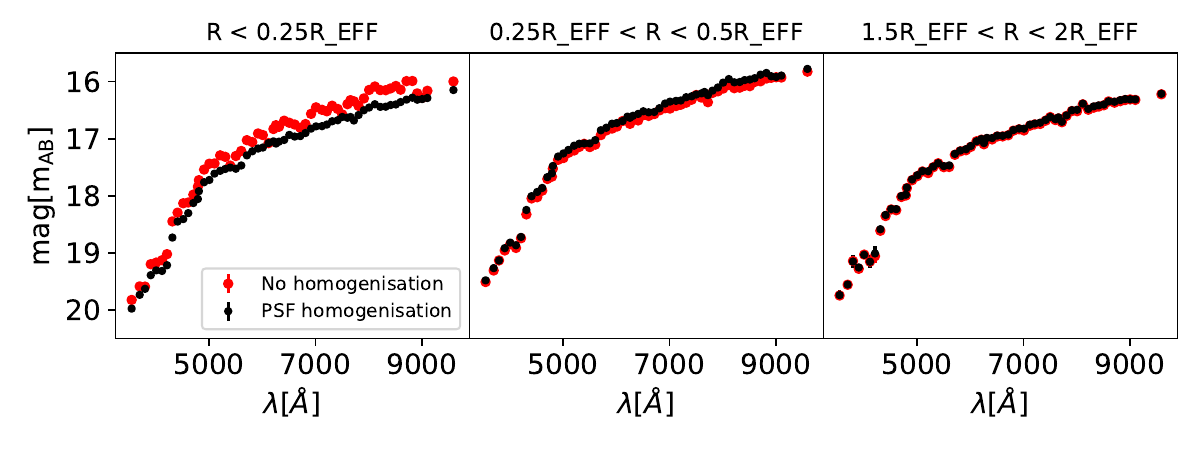}
    \caption{Flux comparison of the \js \ of different apertures of the galaxy $2470-10239$ before and after applying the PSF homogenisation for different apertures, from left to right: R$<0.25$~\texttt{R\_EFF}, $0.25$~\texttt{R\_EFF}$<$R$<0.5$~\texttt{R\_EFF}, and $1.5$~\texttt{R\_EFF}$<$R$<2$~\texttt{R\_EFF}. Red points represent the \js \ obtained with the images prior to PSF homogenisation. Black point represent the \js \ after all the images have been degraded to the worst PSF.}
    \label{fig:degradationjs}
\end{figure*}

To study the effect of the homogenisation on  photometry, we compare the \js \ retrieved at different apertures and annuli for the galaxy, both before and after PSF homogenisation (see Fig.~\ref{fig:degradationjs}). The non-homogenised J-spectrum of the innermost aperture shows variations in the bands that are not associated with absorption or molecular bands. These variations either disappear or become significantly smoother after the PSF homogenisation, at the cost of redistributing flux across contiguous apertures. Since the mass-to-light ratio of the models is fixed by construction \citep[see e.g.,][]{Conroy2013}, this flux loss could lead to a lower value for the stellar mass in the innermost aperture. However, the homogenisation also affects the colour of the galaxy: red bands are, in general, more affected by the homogenisation than blue bands. Figure~5 from \cite{Bonoli2020}, shows the average FWHM of the seeing in each filter.  For filters between 4000 and 8000~\AA, the mean seeing remains close to 1 arcsec (with the exception of some individual filters). However, there is a noticeable increase in the FWHM of the seeing from 800 nm. This increase is due to the time of the observations carried out using these filters, when the EGS/AEGIS field reached its lowest elevations. Consequently,  the FWHMs are larger than those that would have been obtained at lower zenith distances with the same atmospheric conditions.

Many studies have pointed out that working with images where the PSF differences have not been taken into account introduces undesired colour biases \citep[see e.g.,][]{Cyprian2010,GonzalezPerez2011,Er2018,Liao2023}. Our SED-fitting analysis heavily depends on the colour of the galaxy, and the non-physical fluctuations of the \js \  may lead to incorrect results or poorer stellar population fitting. Thus, it is preferable for our analysis to use the homogenised \js, despite  the potential underestimation of the stellar mass. In fact, the relation of the mass-to-light ratio and the galaxy colour is well known, and it has also been used to derive stellar masses \citep[see e.g.,][]{Tinsley1981,Bell2001, Bell2003, Gallazzi2009, Ruben2017,LopezSanJuan2019ML}. Therefore, an incorrect galaxy colour would also lead to an incorrect mass-to-light ratio, resulting in an incorrect estimation of the mass.

Another effect visible in Fig.~\ref{fig:degradationjs}, is that the differences between the homogenised and non-homogenised \js \ become less significant with galactocentric distance. This can be expected, since the homogenisation spreads the light of the image. Brightest apertures lose more flux compared to dimmer apertures, since these dimmer apertures ``recover'' flux from other nearby apertures with a similar brightness. Due to the typical Sérsic profile of the surface brightness \citep{Sersic1963,Sersic1968}, the innermost parts of the galaxy are noticeably brighter and the ``recovered'' flux from other apertures is insufficient to account for the lost flux. We can actually see this effect in the second panel. This close aperture to the centre shows a slightly brighter \js \ after homogenisation, due to the spread light from the centre. Consequently, we can expect flatter radial profiles for the properties of the galaxies after homogenisation.

\subsection{Photometry measurements}

We now apply our methodology to the selected spatially resolved galaxies from \mjp \ and study the \js \ and photometry measurements obtained from them. We start by analysing the importance of the background subtraction and then we compare our measurements with the values found in the catalogues for the circular apertures, as well as for \magauto \ and \magpetro.

\subsubsection{Impact of the background subtraction} 
\begin{figure*}
    \centering
    \includegraphics[width=\textwidth]{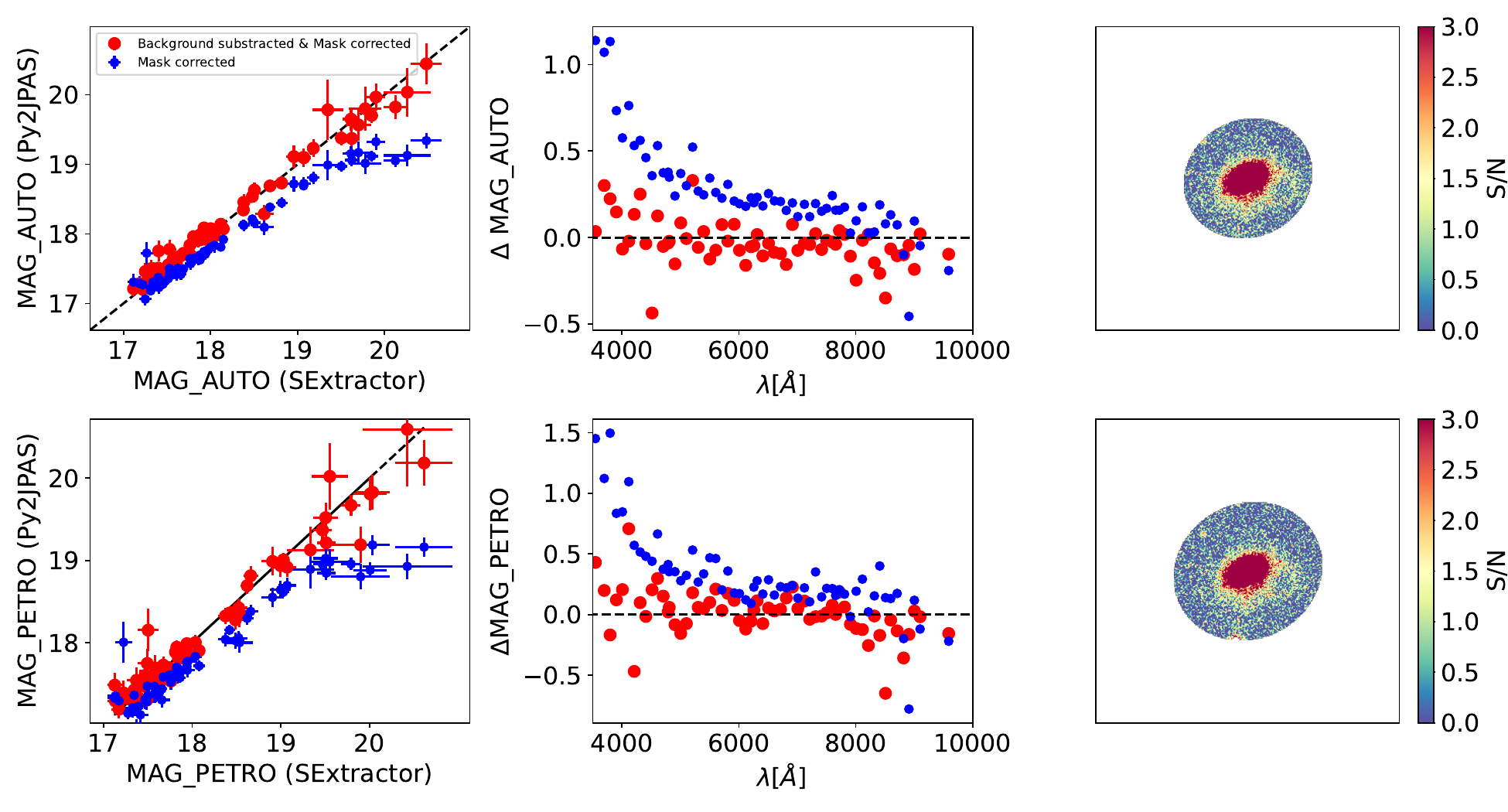}
    \caption[Comparison of the elliptical photometry available in the catalogue and the values retrieved with \PyDJ \ for the galaxy 2241--7608, using the raw image without accounting for masked pixels, the raw image accounting for masked pixels, and the raw image correcting for masked pixels and subtracting the backgroud. ]{Comparison of the elliptical photometry available in the catalogue and the values retrieved with \PyDJ \ for the galaxy 2241--7608, before and after background correction. Blue points represent the values obtained when no background subtraction is applied. Red points represent the values of the photometry obtained applying a background correction. Left panels show the one-to-one relation. The black dashed line shows the one-to-one relation. Middle panels show the difference between our calculations and the \sext \ catalogue values, calculated as $\Delta \mathrm{MAG} = \mathrm{MAG}_{\sext} -\mathrm{MAG}_{\PyDJ} $, as a function of wavelength. The black dashed line shows the 0 target value. Right panels show the S/N image with the aperture used for each calculation.  Top row represents the comparison for \magauto. Bottom row represents the comparison for \magpetro. }
    \label{fig:badbackellPhot}
\end{figure*}

To illustrate the effect of local background estimation, we select galaxy 2241--7608 from our sample and calculate the \magauto \ and \magpetro \ photometry with and without the background correction. We use this galaxy for this case as it is a fainter galaxy than 2470--10239  with a large aperture, containing a significant number of background pixels. This provides a better test for the background correction than galaxy 2470-10239, which is very bright and where the background contribution should be negligible. The comparison can be seen in Fig.~\ref{fig:badbackellPhot}. The differences between the calculated values and the values from the catalogue are  significant when the background is not taken into account, reaching offsets greater than 1~mag. However, the background subtraction greatly improves the comparison for both \magauto \ and \magpetro, with most values within a difference of $0.1$~mag. This correction is more important in bluer bands, which are also dimmer than redder bands. 

However, brighter galaxies show good agreement between our values and those in the catalogues, without the need for a background correction. After inspecting the galaxies individually, we observe a similar trend: galaxies whose apertures contain mostly pixels with a high S/N ratio (close to or greater than 3 in the \rb{} band) do not require background correction to match the catalogue values, while galaxies with a significant fraction of low S/N pixels require background correction. This can be expected, as the contribution of the background to high S/N pixels (usually bright pixels) is likely negligible, while low S/N pixels (usually dimmer pixels) are more affected by background correction. We choose to apply this correction to every aperture in every segmentation, as it can be crucial for the accurate measurement of magnitudes in outer and/or dimmer apertures.

\subsubsection{Photometry comparison}

\begin{figure*}
     \centering
     \resizebox{\textwidth}{!}{%
      \includegraphics{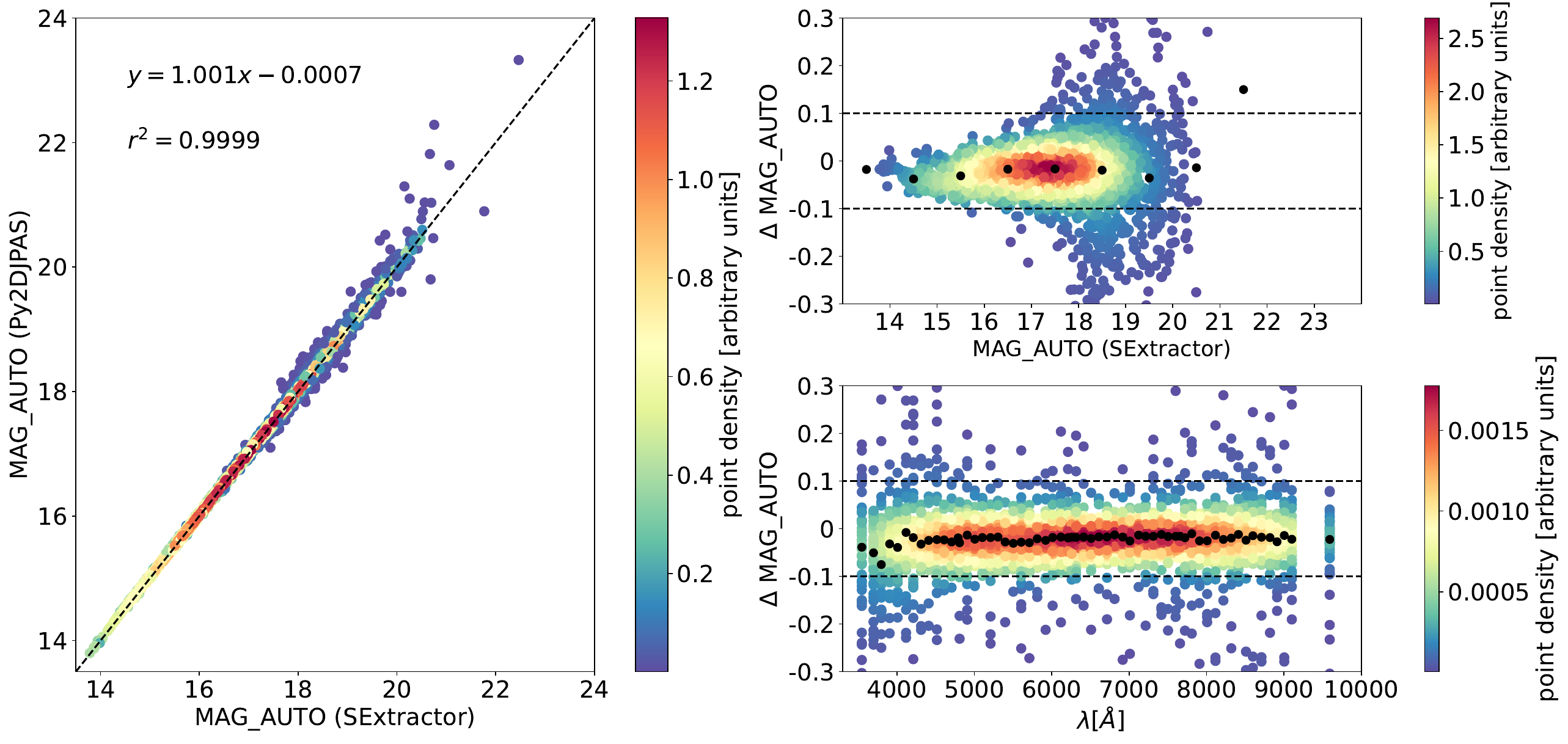}}
     \caption[Comparison of the \magauto \ photometry of the catalogue and the one obtained with our methodology]{Comparison of the \magauto \ photometry of the catalogue and the one obtained with our methodology. Left panel shows the one-to-one relation. Upper right panel shows the difference of the magnitudes (\sext - \PyDJ) for each filter for each galaxy as a function of the magnitude of the band. Bottom right panel shows the difference for each filter for each galaxy as a function of the pivot wavelength of the filter. Colour scale represents the density of points. Black points represent the median value in each brightness bin and wavelength bin.}
     \label{fig:MAGAUTOdensity_corr}
\end{figure*}

For elliptical photometry, we use \magauto \ and \magpetro \ for our tests. We use the non PSF-homogenised images since, according to \cite{Bonoli2020}, the SExtractor photometry was computed from these images. For elliptical photometry, we use \magauto \ and \magpetro \ for our tests. These apertures aim to estimate the complete ($\sim90~\%$) flux of the galaxy, providing the flux within a Kron-like and Petrosian-like aperture.

We first compare the results obtained for \magauto \ (see Fig.~\ref{fig:MAGAUTOdensity_corr}). We observe a tight correlation around the one-to-one relation, with most of the points lying perfectly along this relation. There are a few scattered points that show a larger offset. After individual inspection, we identified these points as corresponding to a couple of fainter galaxies, where either a brighter source was not perfectly masked or the background subtraction needs improvement. In fact, the values of the fit show an almost perfect 1:1 relation, with a slope of $1.001$, an intercept value of $-0.0007$ and a coefficient of determination $r^2 =0.9999$.   Regarding the offsets by brightness, we find that the largest ones occur at magnitudes fainter than 18~mag. However, those points correspond only to a few galaxies. Additionally, the median difference is of the order of $\sim 3$~\% and shows no significant bias.When studied by wavelength, we find that none of the filters show significantly larger offsets than the others, and the median difference remains around $\sim 3$~\% for all the filters. In both cases, the vast majority of points lie within the $[-0.1, 0.1]$~magnitude range or have a relative error below 10~\%. The median values are close to 0 ($\sim 0.03$), regardless of the wavelength or the \rb{} magnitude. This graph shows that, even though some values exhibit some larger offsets (of up to $\sim 0.3$~mag), from a statistical perspective our reconstruction is solid. We have also included the mean, median, standard deviation, and $10^{th}$ and $90^{th}$ percentile values of the difference in Tables~\ref{table:AUTO_bright}, \ref{table:AUTO_wl}, and \ref{table:delta_phot}. The values in these tables show that the differences are generally below the error of the ZP ($0.04$), supporting the statistical reliability of our methodology.

The comparison of the \magpetro \ photometry yields similar results (see Fig.~\ref{fig:MAGPETROdensity_corr}). We observe a larger dispersion than in the case of \magauto \ and a slightly worse agreement in the results, although the fit still closely follows the 1:1 relation (a slope of $1.001$, an intercept value of $-0.0007$ and a coefficient of determination $r^2 =0.9999$). Most of the points are still within the $[-0.1,0.1]$~mag range (relative error below 10\%), but we find a few more galaxies than before in ranges with larger differences. Noticeable outliers are more prevalent, which is expected, since the  apertures are larger. Consequently, it is more likely to include improperly masked sources, and there are more dimmer pixels which are more sensitive to the background correction. In fact, the galaxy with the largest systematic deviation from the one-to-one relation (line of points with \magpetro \ between $\sim 16$ and $\sim 18$) is a single galaxy with a very large aperture, which contains some unmasked sources and many background pixels. Regarding the dependence on the brightness and wavelength, we also find that the median values are close to 0, without a significant bias, and differences generally below the error of the ZP ($0.04$, see Tables~\ref{table:PETRO_wl}, \ref{table:PETRO_bright}, and \ref{table:delta_phot}).

We conclude by noting that the comparison with the circular photometry shows an even better agreement than before (see Fig.~\ref{fig:circulardensity}), and the relation becomes tighter at increasing apertures. Nonetheless, all points remain very close to the one-to-one relation. The results from the fitting are all very close to the 1:1 relation, and the difference are again below the error of the ZP (see Table~\ref{table:delta_phot}). This is also a strong proof of concept, since the sizes of the apertures are unambiguously defined, which guarantees that the integrated apertures are the same for both codes. However, we emphasize that this serves as both a sanity check and a test of the overall calibration process. However, these are not the apertures we are truly interested in; rather, our focus is on photometry in apertures that allow us to obtain the radial profiles of the galaxies.

\subsection{Spatially resolved properties: R\_EFF vs AUTO aperture}

\begin{figure}
   \centering
    \includegraphics[width = 0.45\textwidth]{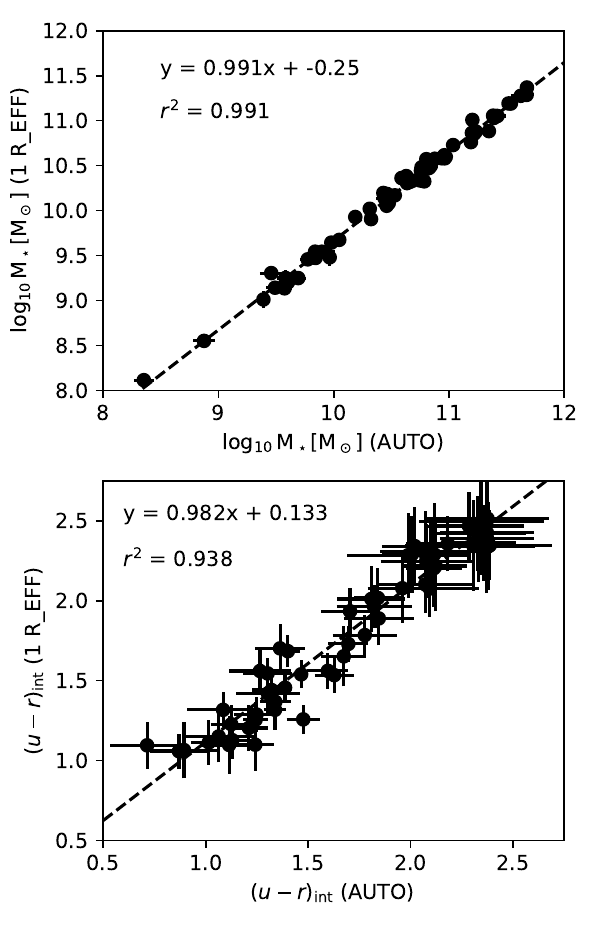}
    \caption{Comparison of properties at 1~\texttt{R\_EFF} and the \magauto \ photometry for the complete galaxy sample. Top panel shows the stellar mass. Bottom panel shows the $(u-r)_\mathrm{int}$ colour. }
    \label{fig:auto_1REFF}
\end{figure}

Another validation of our data and methodology comes from comparing the mass and colour calculated using apertures equivalent to those used for \magauto \ and apertures of 1~\texttt{R\_EFF} (see Fig.~\ref{fig:auto_1REFF}). The relation found for the stellar mass is linear, with very tight coefficient of determination $(r^2  = 0.991)$ and a slope very close to one ($0.991$). This means that, since \magauto \ aims at estimating the total flux of the galaxy, the total mass of the galaxy can be estimated with enough precision by simply scaling the measured value of the mass calculated within an aperture of 1~\texttt{R\_EFF}.

Similarly, the $(u-r)_\mathrm{int}$ colour shows a linear relation with a slightly higher, but still reliable, dispersion $(r^2  = 0.938)$. The slope is not so close to the identity value, but it still indicates that the global colour of the galaxy can be determined scaling the measured value of $(u-r)_\mathrm{int}$ at 1~\texttt{R\_EFF}. On the one hand, this could indicate that measuring the flux of galaxies at 1~\texttt{R\_EFF} is enough to determine their general (integrated) properties, due to the relation of the mass and colour with them \citep[see e.g.][]{Luis2019, Luis2019c, Rosa2021, Julio2022}. On the other hand, it also raises a greater interest in the outer regions of galaxies, since the integrated properties seem to be dominated by inner apertures, but the populations and properties in the outskirts can contain important information on the evolution of the galaxy \citep[see e.g.][]{Coenda2019}.

\section{Discussion} \label{sec:discuss}
In this paper we have demonstrated that we are able to accurately reproduce the values of the photometry  for different apertures in the \mjp \ catalogues. While this is not sufficient on its own for our final goal, which is to obtain the properties of the spatially-resolved galaxies in \mjp, it is a necessary condition. We have shown that this initial requirement has been successfully met.

In this section, we discuss the implications of our methodology by analyzing the SED-fitting residuals of the \js\ apertures for the complete sample. We further address the validity of our approach and the IFU-like capabilities of the \jp\ survey by means of the galaxy 2470--10239as a case study, alongside the spatially resolved photometry for the galaxy sample.

\subsection{Residuals of the SED-fitting}
\begin{figure}
    \centering
    \includegraphics[width=0.5\textwidth]{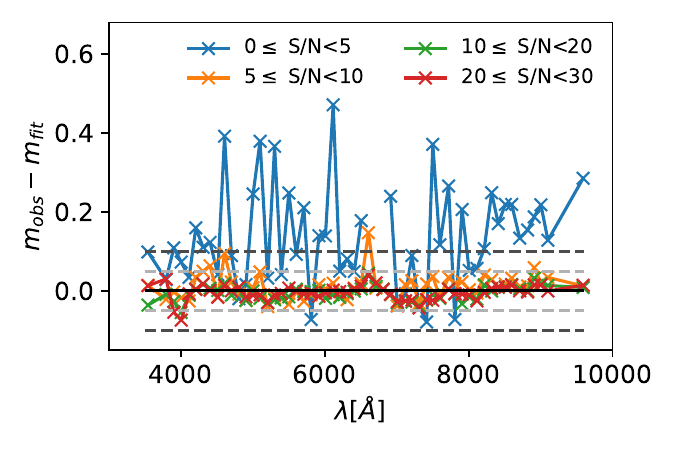}
    \caption[Residuals using rings of the same size as the worst PSF]{Residuals using elliptical apertures that are a multiple of the FWHM of the worst PSF, divided by S/N. Different colours represent different different bins of S/N. Solid black lines represent the null value of residuals. Grey dashed lines in top panels represent the limits $[-0.05,0.05]$ of the residuals. Black dashed lines in the top panels represent the limits $[-0.1,0.1]$ of the residuals. }
    \label{fig:residual_MAX_RES}
\end{figure}

For the complete selected sample, we calculate the residuals for each aperture obtained. These apertures are elliptical annuli whose inner and outer semimajor axis are an integer multiple of the FWHM of the worst PSF. The value of $m_\mathrm{fit}$ is obtained using \baysea \ \citep[see Sect.~\ref{sec:fit_ANN} and][]{Rosa2021}. We study the  median for each filter, using bins of S/N (see Fig.~\ref{fig:residual_MAX_RES}). Low residual values indicate that the code was able to retrieve a combination of stellar populations that resembles the extracted spectra. This serves as a strong validation of our photometry, as obtaining physically plausible spectra suggests that the extraction process is yielding explainable results based on our current understanding of stellar models.

Filters with a high S/N exhibit residuals within the $[-0.05,0.05]$ range, corresponding to a relative flux difference of approximately 5~\%. Most importantly, the residuals do not show a significant bias with wavelength (see Fig.~\ref{fig:residual_MAX_RES}). This difference is on the order of the instrumental error \citep[e.g., $\sigma_\mathrm{ZP} = 0.04$, see][]{Lopez-sanJuan2019} and discrepancies attributed to uncertainties that are not accounted in stellar population models.

However, when the S/N falls below five, the residuals increase significantly, reaching values around $\sim 0.2$ (a equivalent to a relative flux difference greater than 20~\%) and are biased towards positive values for all filters. This suggests that the photometry may be biased because we have probably reached the surface brightness detection limit. To mitigate this, we mask out bands with $S/N<5$ during the SED-fitting. Additionally, in our analysis, we only consider those apertures with a median $S/N > 5$ in the bluer filters ($\lambda_\mathrm{pivot} < 5000$~\AA), since these filters contain the so-called 4000-break, which is known to be highly correlated with the stellar population properties of galaxies \citep[see e.g.][]{Rosa2005}. This criterion criterion establishes an effective upper limit for the maximum galactocentric distance at which we can study the galaxy.

\subsection{IFU-like Capabilities of J-PAS}
The instrumental characteristics of JPCam and its photometric systems make J-PAS an ideal survey for IFU-like studies, as it provides spatially resolved data (2D pixel-by-pixel) combined with spectral information through its narrow-band filter system. This approach is equivalent to data cubes obtained from IFU instruments with a low spectral resolution ($R\sim 60$), allowing the study of the spatially resolved properties of galaxies across a broad wavelength range.

This capability is particularly valuable for the study of bright galaxies in the Local Volume, which often have angular diameters of several arcminutes. Such galaxies cannot be fully observed with conventional IFU instruments, which typically have a field of view (FoV) of $\sim 1$~arcmin or less. The ability of J-PAS to map such extended structures makes it a unique and powerful tool for investigating the stellar populations, and ionized gas properties of nearby galaxies.

Moreover, the large FoV of JPCam enables the efficient study of entire galaxy clusters, such as the Hercules or Coma clusters, in a single pointing. This allows for spatially resolved studies of all member galaxies within these clusters, facilitating research into environmental effects on galaxy evolution, such as ram pressure stripping, mergers, and interactions. The combination of wide-area coverage and IFU-like spectral information opens new possibilities for statistical studies of spatially resolved galaxy properties in different cosmic environments, bridging the gap between large spectroscopic surveys and deep integral field observations.

J-PAS will also be highly competitive for IFU-like studies of extended galaxy samples at $z \leq 0.1$. This redshift limit corresponds to spatial scales of $\leq 1.8$~kpc/arcsec, allowing spatial sampling of galaxy properties at a resolution of $\sim 0.5$~HLR for the most distant galaxies, assuming a typical HLR of 4~kpc and a seeing of 1~arcsec. Even better spatial sampling will be achievable for closer galaxies. Furthermore, J-PAS will be highly competitive when compared to MaNGA, which observed $\sim 10,000$ galaxies at $z \leq 0.15$, selected primarily by their stellar mass in the range of $10^{10}$ to several times $10^{11}$~M$_\odot$. MaNGA employs hexagonal fiber bundles that, for many galaxies, only cover the central 1~HLR, with individual fibers having a size of $\sim 2$~arcsec. It is worth mentioning that, for a single galaxy, J-PAS will need to observe 56 times in order to obtain the complete \js, which makes IFU surveys more efficient for the observation of single galaxies. However, thanks to its FoV, as well as its observation strategy, can observe numerous galaxies in different filters in a single exposure, effectively increasing the number of galaxies observed with all the filters as the observed footprint expands. It is important to take into account that the survey was designed to be efficient towards large areas, rather than single areas. Furthermore, given its pixel scale, the spatial resolution of J-PAS will be determined by the seeing conditions of the observations, which are expected to be smaller than the typical fiber size in IFU surveys.
J-PAS, on the other hand, offers a significant advantage over MaNGA, as it will be able to extract spatially resolved properties (including stellar population properties and ionized gas) out to 2–4~HLR, with a resolution of $0.23$~arcsec/pixel (or $0.5$~arcsec/pixel if a 2×2 binning is applied in J-PAS data). This capability will guarantee a simultaneous study of both bulge and disk properties, making J-PAS a powerful tool for understanding the structural and evolutionary processes of galaxies.

As a closing point of this discussion, we present the analysis of the galaxy 2470-10239, using the available data from the MaNGA survey for this galaxy allows for comparing of the performance of our code applied to the \mjp \ data with the results obtained using the MaNGA data, in a similar way to the example shown in \cite{Bonoli2020}.

\begin{figure*}
    \centering
    \includegraphics[width = 0.9 \textwidth]{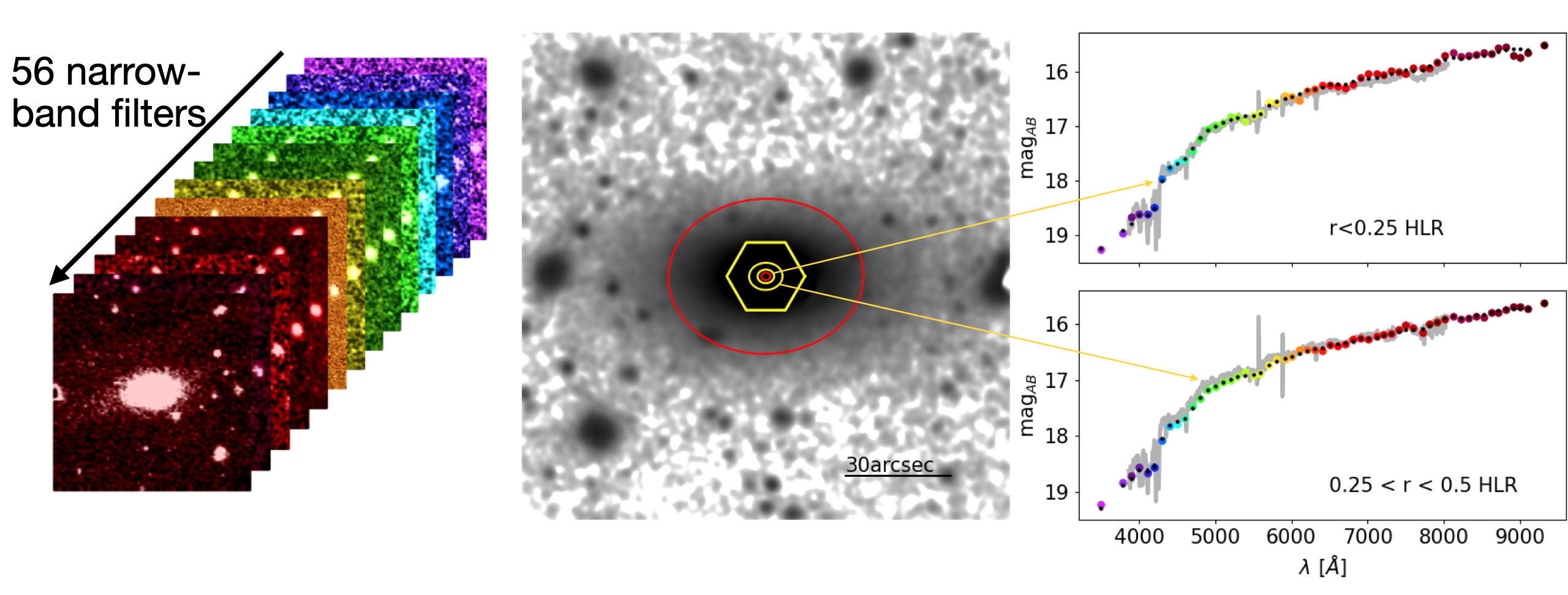}
    \caption{IFU-like capabilities of the J-PAS data. First panel: representation of the different filter images contained in a data cube from \mjp. The colour of each slide represents the band. Second panel: \rb \ image of the galaxy 2470--10239. Innermost red ellipse shows the FWHM of the worst PSF. Outermost red ellipse shows the distance at which  \js \ with a median $S/N>5$ in the filters with $\lambda_{\mathrm{pivot}} <5000$~\AA \ can be obtain, using elliptical annuli, increasing their semimajor axis in steps of the FWHM of the worst PSF. The yellow hexagon shows the FoV of the larges MaNGA IFU. Yellow ellipses show the matching apertures with MaNGA data, using the HLR calculated by \pycasso \ on the MaNGA data cube ($r<0.25$~HLR and $0.25< r <0.5$~HLR). Right panel shows the comparison of the \mjp \ and MaNGA data in those apertures. Grey lines represent the MaNGA spectroscopic data. Colour points represent the \mjp data. Black points represent the SED-fitting obtained with \baysea.}
    \label{fig:artisticfigure}
\end{figure*}

As previously mentioned, this kind of data cubes offered by J-PAS (see Fig.~\ref{fig:artisticfigure}) are actually an excellent IFU-like resource, with a very large number of filters spanning the complete optical wavelength range. Additionally, the contiguous footprint allows us to study large galaxies at very distant galacto-centric radii. To demonstrate this quantitatively, we show the flux image of the \rb{} band at the maximum distance at which  \js \ gets a $S/N > 5$ in the bluer filters ($\lambda < 5000$~\AA, see Fig.~\ref{fig:artisticfigure}) can be retrieved. All this when using the maximum spatial resolution allowed by the PSF. The distance at which this is achieved is greater than 4~\texttt{R\_EFF}, even surpassing the distances reached by other IFU surveys, as shown by the comparison with the FoV of IFU used in MaNGA to observe the same galaxy (see Fig.~\ref{fig:artisticfigure}). This capability is especially valuable for studying the outermost regions of galaxies, which are more susceptible to environmental processes, such as ram preassure stripping \citep[see e.g.][ and references therein]{Coenda2019}. Furthermore, this capability also offers the possibility to study the low surface brightness regions of galaxies (see Eskandarlou, in prep. for an example using the \jp \ for the study of the low surface brightness regime).

We also compare the magnitudes obtained from MaNGA data and \pycasso \ with those derived from \PyDJ \ for the \mjp \ data for equivalent apertures, similarly to the result presented in \cite{Bonoli2020}, using two different extractions, an elliptical aperture of $r<0.25$~HLR and an elliptical ring of $0.25< r <0.5$~HLR (see Fig.~\ref{fig:artisticfigure}). This HLR corresponds to the half light radius calculated by \pycasso \ on the MaNGA data cube. The extractions obtained using \PyDJ \ show a very good agreement over all the spectral range with  the spectroscopic data. This serves as a sanity check in several respects. Firstly, this result implies that \PyDJ \ is able to retrieve fluxes and magnitudes consistent not only with \sext \ and the \mjp \ catalogues, but also with other surveys and codes. Secondly, it also shows that these fluxes and magnitudes are consistent not only for integrated values of the galaxy, but also for apertures at different distances from the galactic centre. Lastly, it also shows that the calibration is  consistent with other surveys, not only when studying integrated objects, but also when working with spatially resolved ones.

\begin{figure}
    \centering
    \includegraphics[width=0.45\textwidth]{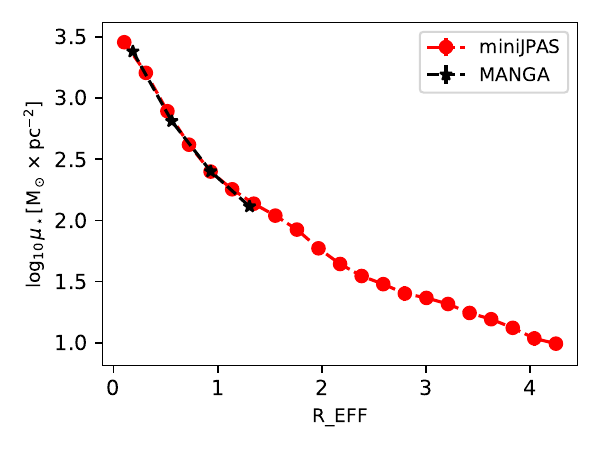}
    \caption{Comparison of the radial profiles of the stellar mass density of the galaxy 2470--10239 obtained with our methodology and with MaNGA data. Red dots represent the profile obtained with \PyDJ \ and \baysea \ using the data from \mjp. Black stars represent results obtained with \pycasso \ and \starlight \ using the data from MaNGA.}
    \label{fig:MANGAradialSPPcomp}
\end{figure}

The comparison of the mass density obtained with \PyDJ \ for the \mjp \ data and the results obtained with \pycasso \ and \starlight \ \citep{cid2005} 
for the MaNGA data can be seen in Fig.~\ref{fig:MANGAradialSPPcomp}. Even using different bins, codes, and data, the mass density profiles fit perfectly. Even though the stellar mass density is very related to the light distribution of the galaxy due to the mass-to-light relation of the stellar models, this is still a very good proof of the validity of our methodology.

The stellar mass density shows a very similar profile to that found by \cite{Rosa2014, Rosa2015, Bluck2020} and \cite{Abdurro2023} for elliptical and quenched galaxies. That is, the surface mass density decreases as the distance to the center increases. The profile found resembles a Sérsic profile \citep{Sersic1963} or a de Vaucouleurs profile \citep{deVaucouleurs1948}. Our result is more similar to the profile found by \cite{Rosa2015} than to those found by \cite{Bluck2020} and \cite{Abdurro2023}.  

The values of the stellar mass density range from $\log_{10} \mu_\star \approx 3.5$ in the innermost apertures down to $\log_{10} \mu_\star \approx 1$ in the outermost apertures. The values found for this type of galaxy by \cite{Rosa2014, Rosa2015} range from $\log_{10} \mu_\star \approx 4$ in the innermost apertures to $\log_{10} \mu_\star \approx 2.5$ at $2~R_{\text{EFF}}$ (assuming a Salpeter IMF). At that distance, the stellar mass surface density is $\log_{10} \mu_\star \approx 1.7$ (assuming a Chabrier IMF).  

We note that we are comparing a single galaxy with the average values found by \cite{Rosa2015}. Additionally, the results presented in that work use a Salpeter IMF, which, as also found in that study, on average provides a stellar mass $0.27$~dex larger than the \cite{Chabrier2003} IMF used in our analysis. Therefore, the profiles are compatible, although Our result exhibits a steeper decline in $\log_{10} \mu_\star$ as a function of radius.  

Comparing the values of $\log_{10} \mu_\star$ that we obtain with those shown by \cite{Bluck2020} for quiescent galaxies, the agreement is almost perfect ($\log_{10} \mu_\star \approx 3.5$ in the central apertures, $\log \mu_\star \approx 2.2$ at $1.4~R_{\text{EFF}}$). The surface mass density profiles found by \cite{Abdurro2023} range from $\log_{10} \mu_\star \approx 4$ in the central apertures to $\log_{10} \mu_\star \approx 2$ at $4~R_{\text{EFF}}$. However, they report a dispersion in these values that is consistent with our findings.

\subsection{Local relations: Spatially resolved $\Sigma_\mathrm{SFR}$ and colours}

\begin{figure}
    \centering
    \includegraphics[width=0.5\textwidth]{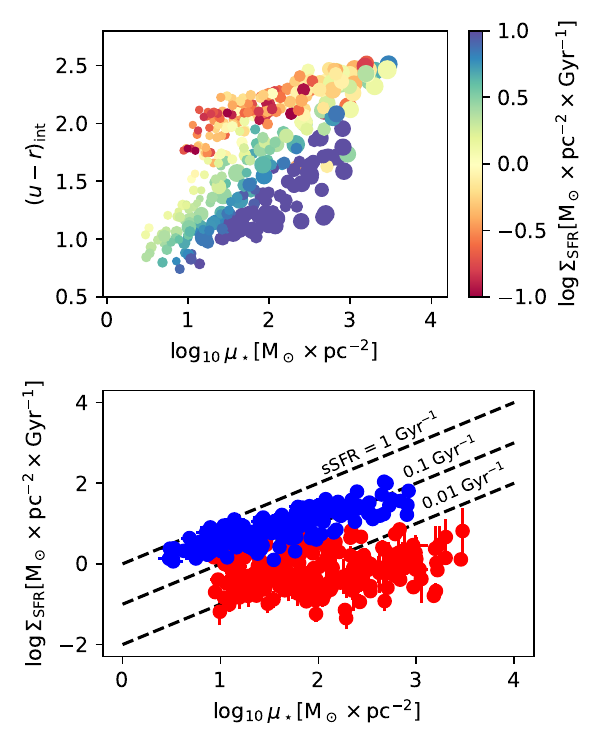}
    \caption{Stellar mass density, colour, and $\Sigma_\mathrm{SFR}$ relation. Top panel: stellar mass surface density--colour diagram, colour coded by $\Sigma_\mathrm{SFR}$. Each point represents the values obtained for an aperture area. Bottom panel: local SFMS. Red points represent red galaxies. BLue points represent blue galaxies. Black dashed lines represent loci of constant sSFR, with sSFR$=0.01$, $0.1$, and 1~Gyr$^{1}$, respectively.}
    \label{fig:SFMS}
\end{figure}

The IFU-like capabilities of J-PAS can also be Effectively utilised to extract other spatially resolved properties of galaxies, such as the intensity of the star formation and their spatially resolved colours. 
Results from the SED fitting also enable us to calculate the intensity of the SFR in the apertures. Using the SFHs of the apertures, we sum the stellar mass formed over the past $10^8$ years and divide it by this time period. Dividing this quantity by the area of each aperture allows us to compute the SFR intensity, $\Sigma_\mathrm{SFR}$. We use this parameter to examine the evolutionary state of the various galaxy regions. The logarithm of the stellar mass surface density ($\mu_\star$) is made dimensionless by using  $\log_{10}(\mu_\star / (M_\odot/\mathrm{pc}^2) $. We classify galaxies into red and blue populations using the selection criterion from \cite{Luis2024}, which is an adaptation of the method proposed by \cite{Luis2019} and previously employed by \cite{Rosa2021} to separate galaxy populations in \mjp\ into red and blue categories. We classify galaxies as red if they satisfy the following condition:
\begin{equation}
    (u-r)_{\mathrm{int}} > 0.16 \left(\log_{10}( M_{\star}) - 10\right) - 0.254 \times (z - 0.1) + 1.689,
\end{equation}
where $(u-r)_{\mathrm{int}}$ is the intrinsic $(u-r)$ colour of the galaxy, corrected from extinction, and $M_{\star}$ is the total stellar mass of the galaxy, both calculated from SED-fitting.

There is a clear relationship between $\Sigma_\mathrm{SFR}$, the colour, and the mass density of the galaxies in our sample (see the top panel of Fig.~\ref{fig:SFMS}). This diagram shows a higher $\Sigma_\mathrm{SFR}$ in apertures with bluer colours, while apertures with a lower $\Sigma_\mathrm{SFR}$ tend to be redder. This trend is analogous to the relationship found in integrated galaxies, where red galaxies generally exhibit lower SFRs than blue galaxies, as demonstrated, for example, in the well-known star formation main sequence (SFMS) \citep[see, e.g.,][]{Brinchmann2004,Noeske2007}. However, $\Sigma_\mathrm{SFR}$ also correlates with the stellar mass surface density, resulting in a crosswise distribution of the parameter in the diagram. Consequently, for a given value of $(u-r)_\mathrm{int}$, we observe a higher value of $\Sigma_\mathrm{SFR}$. Considering the typical profiles of $\mu_\star$ and $\Sigma_\mathrm{SFR}$, where central apertures typically show higher values of these quantities \citep[see, e.g.,][]{Rosa2014, Rosa2016, Rosa2017}, this result is expected.

To further investigate this relationship, we also examine the local SFMS, as explored in other studies \citep[see, e.g.,][]{Sanchez2013, Rosa2016, Cano2019SDSS}. Our results show that apertures within blue galaxies exhibit a clear trend, where $\Sigma_\mathrm{SFR}$ increases with $\mu_\star$, consistent with the literature and with the trend observed in the colour-mass-density diagram. These findings serve as a validation of the J-PAS data's capability to analyse spatially resolved galaxies in a manner consistent with previous results obtained from other surveys. Furthermore, the large sample of galaxies provided by J-PAS, free from pre-selection bias, offers significant advantages. This is particularly valuable for studying the role of the environment in galaxy evolution.

\section{Summary and conclusions}

This work describes and tests \PyDJ, our tool to automatise the complete analysis of the spatially-resolved galaxies of the \mjp \ survey, which will also be useful for the study of this type of galaxies in the \jp \ survey after adapting certain steps. The goal is to provide a methodology that can be used in order to study the properties of galaxies at local scale, taking advantage of the IFU-like capabilities of these surveys. These capabilities are mainly based on the large number of bands of the survey (56 narrowbands plus 4 broadbands), which provide a spectral resolution enough to obtain the stellar population properties of galaxies through SED-fitting. Their large FoV, which  efficiently covers large, contiguous areas of the sky allows for observing large galaxies in their full extent. This FoV  also removes the preselection bias, which is useful for the construction of samples of galaxies belonging to clusters and groups which can be used to study the role of environment on galaxy evolution at local scale. 

The different steps of our methodology can be summarised in: (i) download of scientific images and tables, (ii) masking of the nearby sources (iii) PSF homogenisation, to provide equivalent apertures thought all the filters, (iv) Magnitudes and flux calculations, and (v) SED-fitting and emission line estimation.

The downloaded data include photometric tables (\texttt{MagABDualObj} and \texttt{FLambdaDualObj}), zero-point calibrations, photometric redshifts from JPHOTOZ, and cross-matching with SDSS. Additionally, PSF measurements and scientific images containing ADUs and WCS transformation parameters are obtained. Square image stamps of size 30 times \texttt{A\_WORLD} are used for analysis.

To mask bright star and artifacts, we make use of the masks provided by the J-PAS database. Nearby sources are masked using a custom mask obtained using tools from \texttt{astropy} library. PSF homogenization is performed via Gaussian kernels, ensuring consistent aperture photometry at the expense of spatial resolution. This process effectively removes non-physical features in the \js.

We estimate the contribution of masked pixels in order to improve the measurement of the total flux within the apertures, and local background subtraction is performed using elliptical annuli based on \texttt{A\_WORLD} and \texttt{B\_WORLD}. This approach improves the agreement between \magauto \ and \magpetro \ by up to 1 magnitude for faint galaxies.

Our methodology accurately reproduces the values of the magnitudes in the \magauto, \magpetro, and circular photometry of the publicly available \mjp \ catalogue for our  sample of spatially-resolved galaxies in \mjp\ \citep{Bonoli2020}, with a precision typically around $\sim 0.1$~mag. Additionally, the median difference in each filter is of the order of $\sim 3$~\%.

By performing a SED-fitting of the \js \ obtained  within elliptical rings, with their axes increasing in steps of the FWHM of the worst PSF, we find that the residuals of the fitting are below $\sim 10$~\%, with no significant bias on the wavelength, for apertures with S/N$>5$. This condition can be achieved for distances larger than \texttt{R\_EFF} for the brightest galaxies in our sample, reaching outer regions of these galaxies that would be outside the FoV of other surveys, offering the possibility of studying regions of lower surface brightness or the role of environment at local scale. 

We demonstrate the IFU-like capability of J-PAS by analyzing the nearby galaxy 2470–10239 at \( z = 0.078 \) with Py2DJ-PAS, which was also observed as part of the MaNGA survey. The spatially resolved photometric multiband extractions in two inner apertures of the galaxy show very good agreement with the spectroscopic flux from the MaNGA data cube.  
In particular, the results obtained from the full spectral fitting of MaNGA spectra with \starlight\ and the SED-fitting with BaySeAGal using the J-spectra provide identical profiles for the stellar mass surface density in both approaches. These results support the stability and reliability of the calibrations and methodology.  

Moreover, we demonstrate the capability of J-PAS to derive spatially resolved stellar population properties, such as the stellar mass surface density, out to several \( R_{\text{eff}} \), whereas most of the MaNGA data are restricted to the inner \( 1.5\,R_{\text{eff}} \). Furthermore, we show the capabilities of \texttt{Py2DJPAS} to derive the spatially resolved \(\Sigma_{\text{SFR}} - \log_{10} \mu_\star\) relation, in a similar way to IFU studies of the local main sequence relation.

Throughout this paper, we have shown that our methodology provides robust photometric measurements. This is the first step toward our ultimate goal of studying the spatially resolved properties of a sample of galaxies in nearby groups, which will be addressed in an upcoming paper.

\begin{acknowledgements}
J.E.R.M., L.A.D.G., R.M.G.D., G.M.S., R.G.B., and I.M. acknowledge financial support from the Severo Ochoa grant CEX2021-001131-S funded by MICIU/AEI/ 10.13039/501100011033.
J.E.R.M., L.A.D.G., R.M.G.D., G.M.S., and R.G.B., are also grateful for financial support from the project PID2022-141755NB-I00, and proyect ref. AST22\_00001\_Subp 12 and 11 with fundings from the European Union - NextGenerationEU»; the Ministerio de Ciencia, Innovación y Universidades ; the Plan de Recuperación, Transformación y Resiliencia ; the Consejería de Universidad, Investigación e Innovación fromth Junta de Andalucía and the  Consejo Superior de Investigaciones Científicas. 
I.M. acknowledges financial support from the project PID2022-140871NB-C21.
I.B. has received funding from the European Union's Horizon 2020 research and innovation programme under the Marie Sklodowska-Curie Grant agreement ID n.º 101059532. This project was extended for 6 months by the Franziska Seidl Funding Program of the University of Vienna.
Based on observations made with the JST/T250 telescope and JPCam at the Observatorio Astrofísico de Javalambre (OAJ), in Teruel, owned, managed, and operated by the Centro de Estudios de Física del Cosmos de Aragón (CEFCA). We acknowledge the OAJ Data Processing and Archiving Unit (UPAD) for reducing and calibrating the OAJ data used in this work. Funding for the J-PAS Project has been provided by the Governments of Spain and Aragón through the Fondo de Inversión de Teruel, European FEDER funding and the Spanish Ministry of Science, Innovation and Universities, and by the Brazilian agencies FINEP, FAPESP, FAPERJ and by the National Observatory of Brazil. Additional funding was also provided by the Tartu Observatory and by the J-PAS Chinese Astronomical Consortium.
This paper has gone through internal review by the J-PAS collaboration. We thank Helena Domínguez Sánchez for her work as an internal referee.
\end{acknowledgements}

   \bibliographystyle{aa} 
   \bibliography{main} 

\appendix

\section{\PyDJ \ description} \label{app:code}
\begin{figure*}
    \centering
    \includegraphics[width=\textwidth]{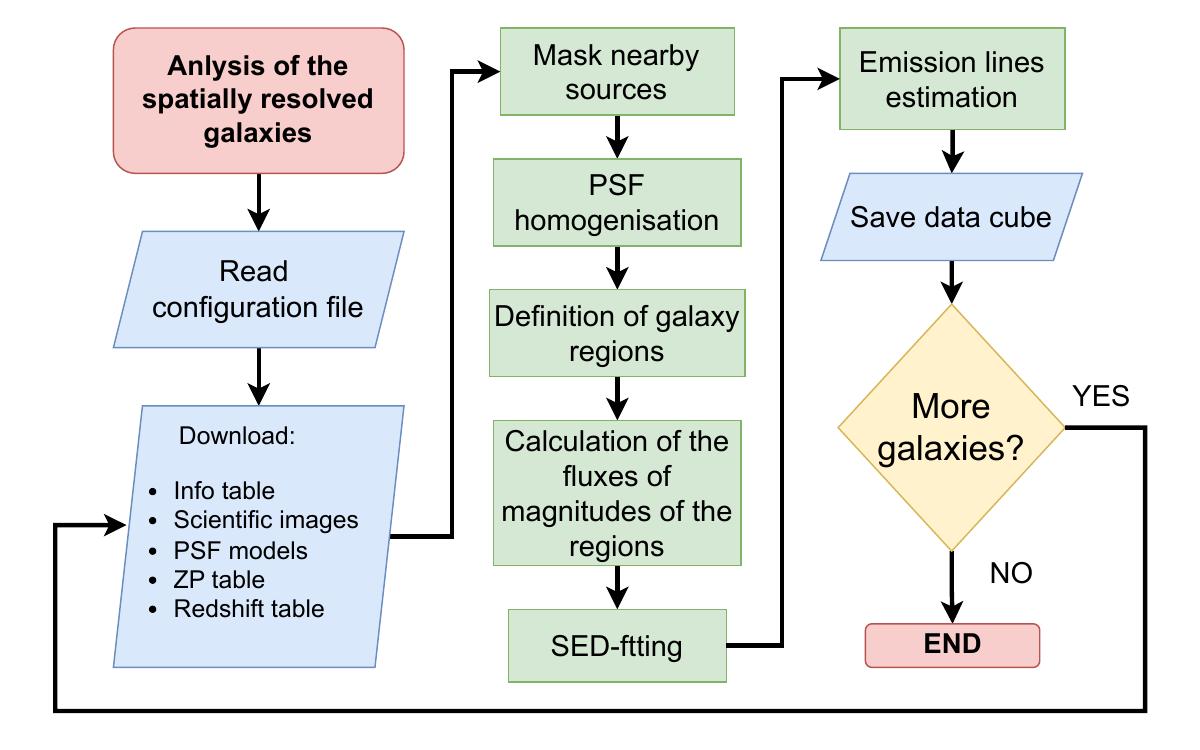}
    \caption[Flow diagram of \PyDJ.]{Flow diagram of \PyDJ. Red ovals Indicate the beginning and end of the different process, blue parallelograms indicate reading and writing input and output files, green rectangles represent computation processes and the yellow diamonds represent decision points. Further details of the processes can be found in Sects.~\ref{sec:code:download}, \ref{sec:code:flux}, \ref{sec:code:masks}, \ref{sec:code:PSF} and \ref{sec:code:photometry}}
    \label{fig:flowchart}
\end{figure*}

In this appendix we explain the main details of \PyDJ, including the libraries used, the values of the required parameters and the reasoning behind certain decisions, as well as the general workflow of the code, summarised in Fig.~\ref{fig:flowchart}.

\subsection{Code initialisation and download of images and tables} \label{sec:code:download}

At the beginning of the process, the code creates a data cube where  all the relevant information for our analysis is saved, mostly scientific tables and images. We first save a table containing relevant information about the photometric filters, mainly the name of the filter and its pivot wavelength. This table is useful to control certain subroutines, keep track of the filters used, as well as to  perform certain calculations, such as the conversion from analog-to-digital units (ADU) into flux units.

The main script then begins the download of the relevant images and tables required for the analysis. All the download process are handled through Astronomical Data Query Language (ADQL) queries. The input from the user are the IDs of the galaxies. 

We then download the  \texttt{MagABDualObj} and the \texttt{FLambdaDualObj} tables for each object. These tables were obtained by running \sext \ on its Dual-Mode using the \rb{} filter as the reference band, and contain useful parameters of each galaxy, such as the different integrated photometry and their errors, the sky coordinates, the ellipse parameters, and the \kron \ and \petror \ radii. We note that the photometry in these tables is given in AB magnitudes and flux in units of $\mathrm{erg} \times \mathrm{s^{-1}} \times \mathrm{cm^{-2}} \times \mathrm{\AA^{-1}}$. 

The next step is the download of scientific images. We download the images for all the filters in order to test the performance of the code in every band. These images are square stamps of 30 times the size of \texttt{A\_WORLD}, that is, the major semi-axis provided by \sext. We find that this size results in an area large enough as to contain the whole galaxy and to estimate the background. We also use the data from the images to create a data cube with shape $(x,y, n_{\mathrm{filters}})$, where  $(x,y)$ are pixels units and $n_{\mathrm{filters}}$ is the number of filters. In the case of \mjp, $n_{\mathrm{filters}} = 60$. We use these cubes to obtain the photometry of the different apertures of the galaxy. Next, we download the PSF measurements provided by the J-PAS database \footnote{\url{https://archive.cefca.es/catalogues/}}, in order to homogenise the PSF of all images. This improves the photometry of inner apertures and provides more reliable data for the SED-fitting analysis (see Section~\ref{sec:code:PSF} for further details).

From the header of the images we use two main type of parameters: those providing information about the World Coordinate System (WCS), that we use to perform pixel-to-sky-coordinate transformations, and the parameters from the image used to calculate the errors of the photometry (\texttt{SNOIFIT}, \texttt{ANOIFIT}, \texttt{BNOIFIT}, and  \texttt{GAIN}; see Sect.~\ref{sec:code:flux}). The  zero points (ZP) of the images and their errors are then downloaded. These are used in Sects.~\ref{sec:code:flux} and \ref{sec:code:photometry} to convert the images from ADU to physical units, as well as to perform the photometry in the pertinent apertures. The photometric calibration is described in \cite{Bonoli2020} and is largely based in the work by \cite{Lopez-sanJuan2019} for \jplus.

The next two items to be downloaded are related to redshift. The first one is the table with  information provided by the $\mathrm{JPHOTOZ}$ package \citep{HC2021}, this is, the \texttt{PhotoZLephare\_updated} table. The last item to be downloaded is the cross correlation table with SDSS, from which we select the spectroscopic redshift (if provided), its error, and its quality flags. We  use this spectroscopic redshift when available, because even though the photo-$z$ provided are in very good agreement with the spectroscopic ones \citep{HC2021}, the precision of the ANN estimations (for instance, the relation between H$\alpha$ and [NII]) improves even further with a more precise value of the redshift \citep[see][for further details]{Gines2021}.

\subsection{Flux conversion} \label{sec:code:flux}
The scientific images provide the counts for each pixel in ADU. In order to convert the ADU into physical units, we use the calibration of the  data \citep{Lopez-sanJuan2019}:
\begin{equation}
    m_{\mathrm{AB}} = -2.5 \log_{10} (ADU) + ZP,
    \label{eq:mag_ab}
\end{equation}
where $ADU$ are the image counts, and $ZP$ is the zero point of the image. Taking Eqs.~6 and 7 from \cite{Logrono2019} for the errors associated to the background and the detector, and assuming Gaussian errors, we obtain that the errors of the magnitudes are:
\begin{equation}
\begin{split}
    \sigma_{m_\mathrm{AB}}^2= \left ( \frac{2.5}{\ln 10}  \right )^2 \times  \\ \left ( \frac {1} {|ADU|  G}    + \frac {S^2_{\mathrm{fit}} N_{\mathrm{pix}} \left (a_{\mathrm{fit}} +b_{\mathrm{fit}} \sqrt{N_{\mathrm{pix}}} \right )^2} {ADU^2}  
      +  \left (  \frac{\ln10}{2.5}  \right ) ^2 (\sigma_{ZP} )^2  \right ) ,
    \end{split}
    \label{eq:mag_err}
\end{equation}
where $G$ is the gain of the detector (in $e^-/ADU$), $N_{\mathrm{pix}}$ is the number of pixels of the integrated aperture, $a_{\mathrm{fit}}$,  $b_{\mathrm{fit}}$ and $s_{\mathrm{fit}}$ are parameters provided in the image headers used to calculate the background noise (\texttt{ANOIFIT}, \texttt{BNOIFIT}, and \texttt{SNOIFIT}, respectively), $\sigma_{ZP}$ is the error of the ZP. In addition, we have taken into account that the images are already background subtracted, so we have used $ADU=C-C_B$, where $C$ stands for the counts of the detector and $C_B$ for the counts of the background, following the work by \cite{Logrono2019}. 

The parameters $a_{\mathrm{fit}}$,  $b_{\mathrm{fit}}$ and $S_{\mathrm{fit}}$ are calculated as part of the data acquisition pipeline. In the forced photometry (dual-mode) using the rSDSS band on the coadded images, the default SExtractor parameters were adopted. For \magauto \ (Kron apertures) and \magpetro \ (Petrosian apertures) the used setting (\texttt{PHOT\_AUTOPARAMS}: \texttt{Kron\_fact}$=2.5$, \texttt{min\_radius}$=3.5$ ,\texttt{PHOT\_PETROPARAMS}: \texttt{Petrosian\_fact}$=2.0$, \texttt{min\_radius}$=3.5$). In this case, the overestimation of errors, resulting from the inclusion of too many background pixels, is a known effect. This issue primarily arises because the interpolation kernel used during the Swarp coaddition introduces correlations among neighbouring pixels. To address this, the image noise is recomputed following a procedure similar to that described in \cite{Labbe2003}. Briefly, for square apertures of area $A$, a series of measurements of the sky flux was performed, varying the side length of these apertures from 2 to 30 pixels. A Gaussian fit was applied to the resulting histogram of sky fluxes, and the sky noise in an aperture covering $A$ pixels was modelled as $\sigma(A)  = s_{\mathrm{fit}} \times \sqrt{A} \times (a_{\mathrm{fit}} + b \times \sqrt{A})$.

In order to account for the error and correlations introduced by the PSF we follow several steps. First, we recompute the values of \texttt{ANOIFIT}, \texttt{BNOIFIT}, and \texttt{SNOIFIT} over the PSF homogenised images. Then, following results by \cite{Klein2021}, we compute the error image of each filter using Eq.~\ref{eq:mag_err} with $N_\mathrm{pix} =1$ for each filter and we convolve it with the squared kernel used for the PSF homogenisation. This gives as a covariance image, $\sigma_\mathrm{conv}$, and we estimate the error of the aperture by using a Gaussian propagation of errors of the pixels in the area, this is:

\begin{equation}
    \sigma_\mathrm{m_\mathrm{AB, PSF} }= \sqrt{\sum_{i.j \in A} \sigma_\mathrm{conv}(i,j)^2}
\end{equation}

\subsection{Masks} \label{sec:code:masks}

Photometric images with a large FoV have several advantages, like unbiased object detection, data that is usually deeper than spectroscopic data (for a same instrument, telescope, and integration time) and the capability to study large objects without aperture bias, among others. However, this large FoV also implies that our targets might be affected by nearby objects, such as stars or other galaxies. Also, images might suffer from artifacts, showing structures that are not real or are out of our interest of study, like cosmic rays, or artificial satellites. Therefore we require a mask, this is, a binary flag for each pixel of the stamp that tells us whether that pixel should be used when calculating the flux or magnitudes of the aperture, or if should not be taken into account, (this is, the pixel should be \textit{masked}). 

With this in mind, we distinguish two types of masks for our work. The first type of masks are computed at complete image level and they are provided in the J-PAS database\footnote{\url{https://archive.cefca.es/catalogues/vo/siap/minijpas-pdr201912/get_global_masks}}. These account for image artifacts and bright stars that bias fluxes and saturate pixels around them, meaning that we cannot use them for our analysis. The use of these masks is advisable, since they account for effects that may not be easy to take into account with other methods. For example, the light of bright stars biases the flux measurement in more pixels than what one could appreciate at simple eyesight, and artifacts may appear only in certain filters.

When studying each galaxy, we want to obtain the flux of the desired apertures and remove pixels that belong to other nearby sources. With this aim,  we develop our own routine to mask these sources. In particular, we use the \texttt{segmentation} module of the \texttt{phoutils} library\footnote{\url{https://photutils.readthedocs.io/en/stable/}}. The module allows us to distinguish between two sub-type of sources: those sources that are in the frame, but well separated from the galaxy, and those sources that might be blended with the target galaxy. This mask depends on two parameters. The first one, $npix$, determines the minimum number of pixels for a detection to be considered as a source. The second one, $rms$, establishes the threshold level to be considered as a detection. We set this threshold parameter as a function of the error background noise for one pixel, this is $threshold = rms \times  S_{\mathrm{fit}}  \left (a_{\mathrm{fit}} +b_{\mathrm{fit}} \right )$.

We use both options in our work. This module was used in other works, such as \cite{galmask2022} but, unlike this work, which aims at providing an  object mask that delimits the target galaxy (in order to be used for morphological studies), we aim at removing the objects that are not the target galaxy. We choose this approach since our goal is to obtain solid and accurate flux and magnitude measurements that provide accurate galaxy properties through SED-fitting and ANN estimation. In this regard, we prefer to be more flexible, and assume that if outermost pixels are still part of the galaxy, these should be included in the aperture and, if they are not part of it, a correct background subtraction should remove their statistical contribution from the integrated flux. This is useful to test the maximum distance at which the properties of the galaxy are still retrievable.

 \begin{figure*}
     \centering
     \includegraphics[width=\textwidth]{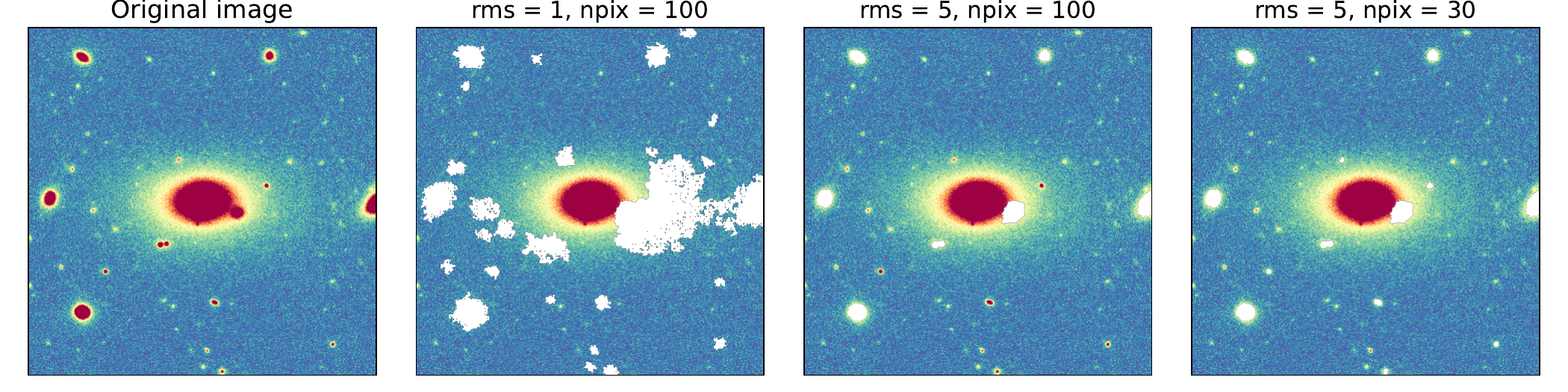}
     \caption[Example of different nearby sources and deblended sources masks obtained for $2470-10239$]{Example of different nearby sources and deblended sources masks obtained for the galaxy $2470-10239$, using different values of the threshold and pixel parameters. From left to right: original image, $rms = 1, npix=100$, $rms = 5, npix=100$, $rms = 5, npix=30$}
     \label{fig:sourcesparam1}
 \end{figure*}

Different values combinations of the parameters that control the nearby source mask can be seen in Fig.~\ref{fig:sourcesparam1}. This Figure shows the compromise reached in our final set, using a value of $rms$ high enough to avoid fake detections, but low enough not as to leave other fainter sources unmasked. It also shows the compromise found for the minimum number of pixels ($npix$), with a value high enough to avoid mixed detections (as in the second panel) at the same time that it removes smaller sources (last panel).

\subsection{PSF homogenisation} \label{sec:code:PSF}
The PSF describes how the light from a point-like source is spread out in an image as a result of the effects of the optics, the atmosphere and the detector. It affects not only point like sources (stars) but also extended ones, such as galaxies. Taking the PSF into account has been proven to be crucial when working with photometric surveys \citep[see e.g.,][]{Heasley1999, Infante-sainz2020, Massari2020}, particularly in multi-band surveys and spatially-resolved studies \citep{Coe2006, Molino2014, SanRoman2018}, since the PSF of the images depends on wavelength and on the different atmospheric conditions of the observations. Therefore, we need to account for these effects in order to obtain reliable measurements of the photometry of the apertures. Otherwise, we might be introducing biases in the colour gradients \citep[see e.g.,][]{Tamura2003, GonzalezPerez2011} or structures that could bias our results. 

In order to avoid these problems, we  homogenise the images to the worst PSF among all the bands for every galaxy. This method has been used in other works such as those by \cite{Loh1986, Labbe2003} or \cite{Kiiveri2021}, since it produces homogeneous apertures in all the bands, at the cost of worsening the spatial resolution. The procedure that we follow is: (i) we parametrize the provided measurement of each PSF assuming a Gaussian approximation, obtaining the FWHM in each filter (ii) we choose the model with the largest FWHM as the worst model, (iii) for each PSF measurement, we compute the kernel that, convolved with the original PSF, would result in the worst PSF, (iv) we convolve each image with its corresponding kernel. This procedure ensures that we use apertures of the same size and that we are not introducing undesired biases and structures on the integration. Otherwise, our measurements would lead to incorrect results and conclusions. 
With this process, we can obtain homogeneous apertures for the images that provide solid magnitude measurements for our galaxies, as we show in Sect.~\ref{sec:results}.

\subsection{Aperture definition and photometry calculation} \label{sec:code:photometry}
The next step  is the definition of the apertures. In this work we use circular apertures/rings and elliptical apertures/rings. The size of these apertures can be defined according to different parameters, such as the \texttt{R\_EFF}, \texttt{A\_WORLD}, or the FWHM of the worst PSF. The number of rings is also flexible. If the elliptical geometry is chosen, in order to calculate the parameters of the ellipse, we first create an elliptical aperture, with semi-major and semi-minor axes $a = 2 \times \texttt{A\_WORLD} \times \texttt{PETRO\_RADIUS}$ and $b= 2 \times \texttt{B\_WORLD}  \times \texttt{PETRO\_RADIUS}$, respectively, in order to limit the influence of external sources. We force the centre of the elliptical aperture to be located in the central pixel of the stamp, in order to avoid an incorrect identification of the source due to brighter pixels of other sources.  Then, we apply a routine from \pycasso \ \citep{Pycasso2017}, which estimates the ellipticity and orientation of the galaxy using the Stokes parameters, as described in \cite{Stoughton2002}.

To calculate the circular photometry, we use the \texttt{SkyCircularAperture} function from the \texttt{photutils} \ library. to define apertures with diameters of $0.8'', 1.0'', 1.2'', 1.5'', 2.0'', 3.0'', 4.0'',$ and $ 6.0''$.  In order to reproduce the \magauto \ and \magpetro \ magnitudes we define equivalent apertures to those used by \sext, using the parameters from the \infot, as well as the \texttt{PHOT\_AUTOPARAMS} chosen for the \mjp \ data reduction \citep[see  Table~C.1 from][]{Bonoli2020}. To define the apertures, we use \texttt{SkyEllipticalAperture} function from the \texttt{photutils} \ library.

In order to improve the measurement of the magnitudes, we also perform a local background estimation and subtraction. For such purpose, we define an ellipse  using the \texttt{SkyEllipticalAnnulus} function. We set as inner semimajor axis $a_{\mathrm{in}}=  4 \times \kron \times \texttt{A\_WORLD}$, and we choose $a_{\mathrm{out}}= 4.5 \times \kron \times \texttt{A\_WORLD}$  as outer semimajor axis, and inner and outer semiminor axes at $b_{\mathrm{in}}=  4 \times \kron \times \texttt{B\_WORLD}$, and we choose $B_{\mathrm{out}}= 4.5 \times \kron \times \texttt{B\_WORLD}$, respectively.  We obtain the average contribution of the background per pixel within this aperture, taking into account the masked pixels. In order to correct the background contribution for each aperture, we simply subtract the product of the average contribution by the number of unmasked pixels to the total counts of the aperture.

We also add an estimation of the counts of the masked pixels within the aperture. For such purpose, we follow a similar approach as in the case of the background, by calculating the average (unmasked) counts in the aperture after subtracting the background and adding this average multiplied by the number of masked pixels to the total counts.

\section{Additional Figures and tables}

In this appendix we include two figures that might be of interest for the reader, but that are less relevant for the main text. The firs one is the flowchart of \PyDJ \ (see Fig.~\ref{fig:flowchart}), where we visually summarise the different steps that the code takes during the analysis, which are explained in Sect.~\ref{sec:method} and tested in Sect.~\ref{sec:results}.

\begin{figure*}
     \centering
     \includegraphics[width=\textwidth]{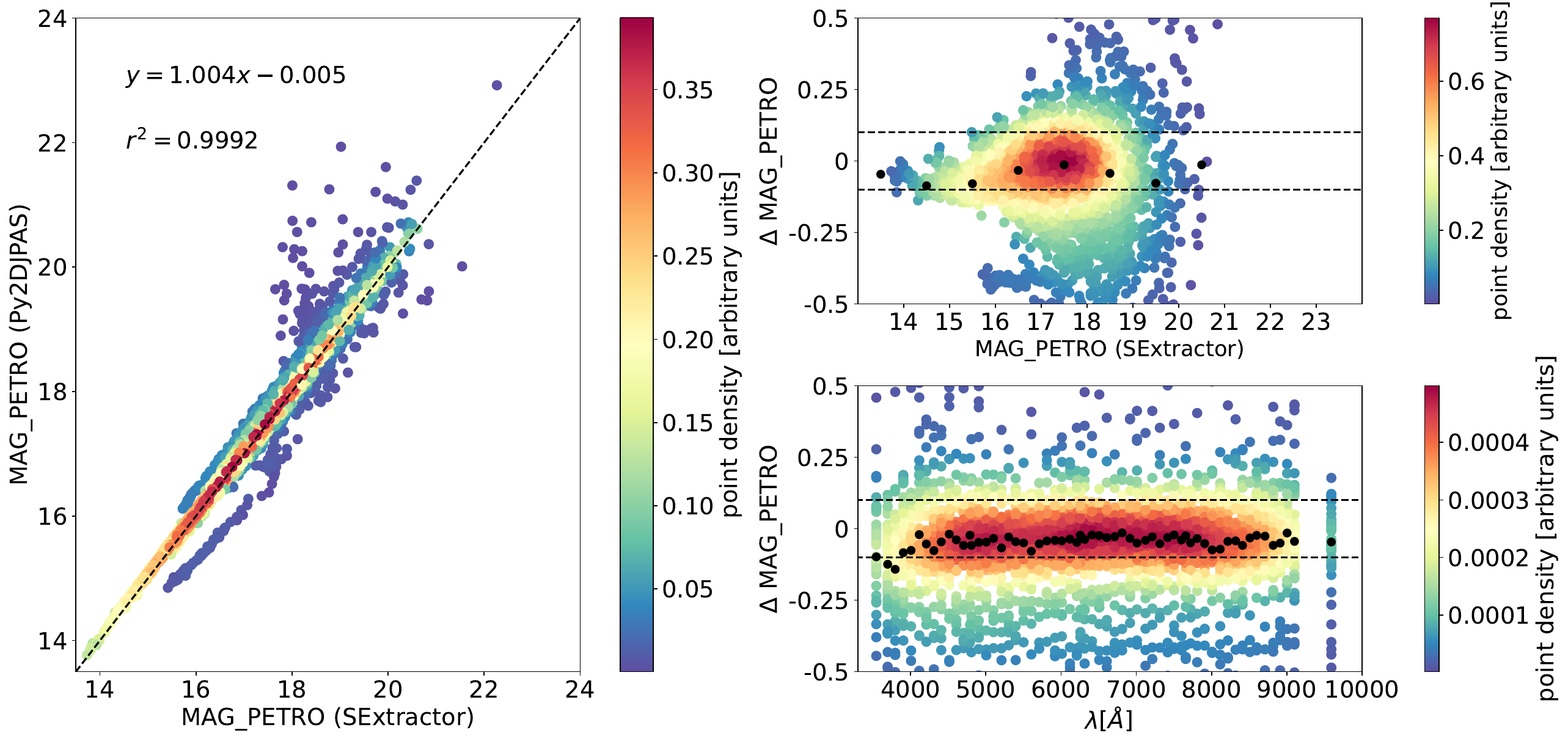}
     \caption[Comparison of the \magpetro \ photometry of the catalogue and the one obtained with our methodology]{Comparison of the \magpetro \ photometry of the catalogue and the one obtained with our methodology. Left panel shows the one-to-one relation. Upper right panel shows the difference of the magnitudes (\sext - \PyDJ) for each filter for each galaxy as a function of the magnitude of the band. Bottom right panel shows the difference for each filter for each galaxy as a function of the pivot wavelength of the filter. Colour scale represents the density of points. Black points represent the median value in each brightness bin and wavelength bin.}
     \label{fig:MAGPETROdensity_corr}
 \end{figure*}

\begin{figure*}
     \centering
     \includegraphics[width=0.8\textwidth]{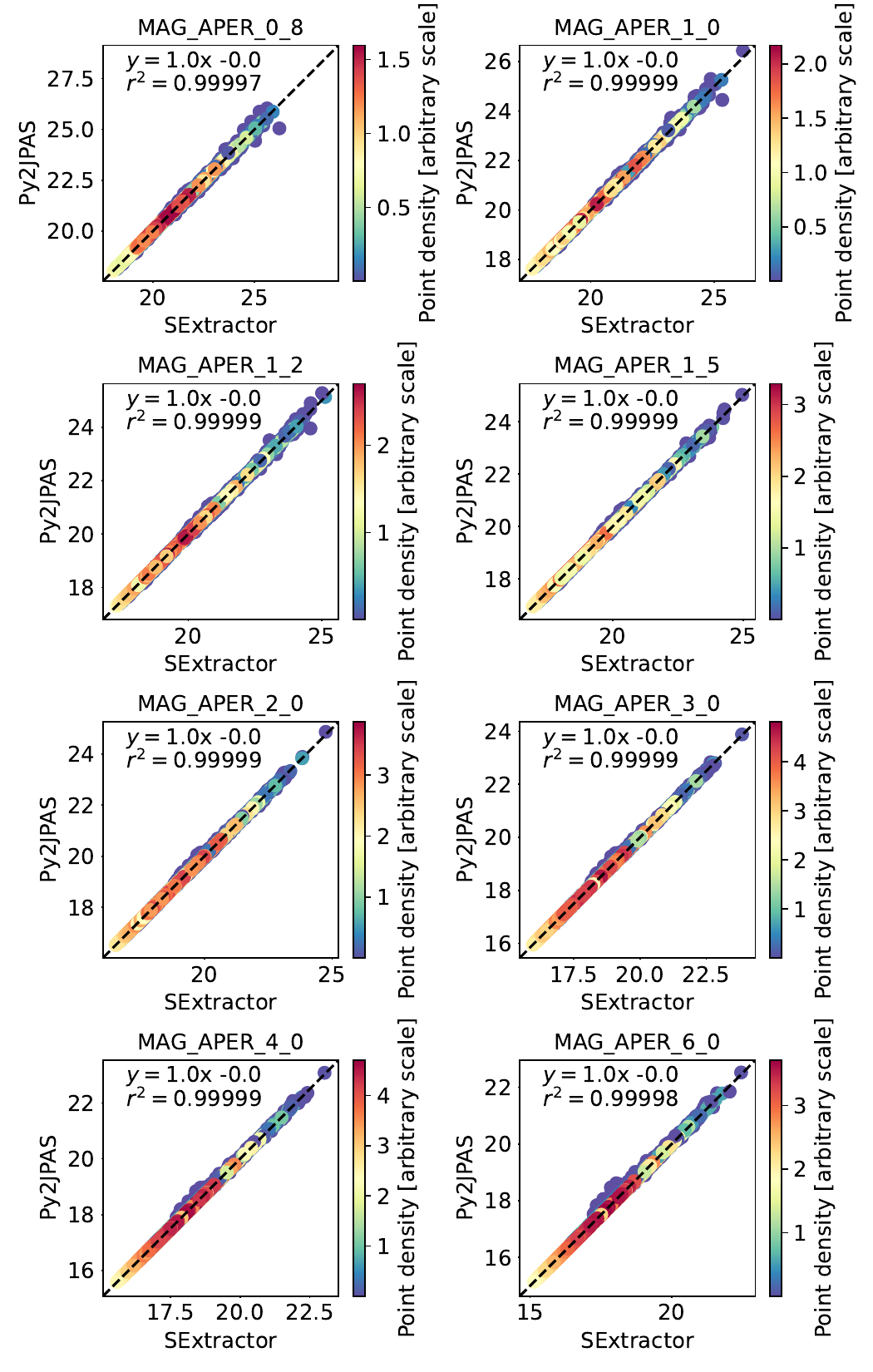}
     \caption[Comparison of the circular photometry of the catalogue and the one obtained with our methodology for the spatially-resolved galaxies in \mjp]{Comparison of the circular photometry of the catalogue and the one obtained with our methodology for the spatially-resolved galaxies in \mjp. Each point represents one filter of one galaxy. Colour scale represents the density of points. The black dashed line represents the one-to-one relation.}
     \label{fig:circulardensity}
 \end{figure*}

We finish by showing the comparison of our measurements with the \magpetro \ phtometry from \sext (see Fig.~\ref{fig:MAGPETROdensity_corr}) and the circular apertures (see Fig.~\ref{fig:circulardensity}), not included in Sect.¢\ref{sec:results} in order to avoid redundancy. This plots show a very good agreement between the two codes, further proving the consistency of our photometric measurements with the values from the \mjp \ catalogues. 

\begin{table*}
\caption{Statistical parameters of the difference $\Delta m_{AB}$ for different brightness bins, using the \magauto \ photometry. The difference is calculated $\Delta \mathrm{MAG} = \mathrm{MAG}_{\sext} -\mathrm{MAG}_{\PyDJ} $. The parameters are, from left to right: mean, median, standard deviation, $10^\mathrm{th}$ percentile, and  $90^\mathrm{th}$ percentile.}             
\label{table:AUTO_bright}      
\centering          
\begin{tabular}{c c c c c c  }     
\hline\hline       
                      
Brightness bin & mean $\Delta m_{AB}$ & median $\Delta m_{AB}$ &  $\sigma(\Delta m_{AB})$ & $P_{10}$ $\Delta m_{AB}$ &  $P_{90}$ $\Delta m_{AB}$\\ 
\hline                    
$13 \le m_{AB} < 14$ & -0.016 & -0.018 & 0.017 & -0.022 & -0.007 \\
$14 \le m_{AB} < 15$ & -0.034 & -0.038 & 0.02 & -0.057 & -0.003 \\
$15 \le m_{AB} < 16$ & -0.029 & -0.031 & 0.028 & -0.064 & 0.011 \\
$16 \le m_{AB} < 17$ & -0.021 & -0.017 & 0.033 & -0.063 & 0.016 \\
$17 \le m_{AB} < 18$ & -0.018 & -0.017 & 0.056 & -0.072 & 0.031 \\
$18 \le m_{AB} < 19$ & -0.025 & -0.019 & 0.101 & -0.146 & 0.079 \\
$19 \le m_{AB} < 20$ & -0.046 & -0.036 & 0.14 & -0.204 & 0.083 \\
$20 \le m_{AB} < 21$ & -0.092 & -0.014 & 0.385 & -0.379 & 0.151 \\
$21 \le m_{AB} < 22$ & 0.15 & 0.15 & 0.722 & -0.428 & 0.728 \\
$22 \le m_{AB} < 23$ & -0.861 & -0.861 & 0.0 & -0.861 & -0.861 \\
\hline                  
\end{tabular}
\end{table*}

\begin{table*}
\caption{Statistical parameters of the difference $\Delta m_{AB}$ for the different filters, using the \magauto \ photometry. The difference is calculated $\Delta \mathrm{MAG} = \mathrm{MAG}_{\sext} -\mathrm{MAG}_{\PyDJ} $. The parameters are, from left to right: mean, median, standard deviation, $10^\mathrm{th}$ percentile, and  $90^\mathrm{th}$ percentile.} 
\label{table:AUTO_wl}      
\centering          
\begin{tabular}{c c c c c c  }      
\hline\hline       
                       
Filter & mean $\Delta m_{AB}$ & median $\Delta m_{AB}$ &  $\sigma(\Delta m_{AB})$ & $P_{10}$ $\Delta m_{AB}$ &  $P_{90}$ $\Delta m_{AB}$\\ 
\hline                    
uJAVA & -0.073 & -0.039 & 0.291 & -0.243 & 0.107 \\
J0378 & -0.102 & -0.075 & 0.161 & -0.222 & 0.023 \\
J0390 & -0.045 & -0.032 & 0.089 & -0.138 & 0.029 \\
J0400 & -0.048 & -0.039 & 0.137 & -0.136 & 0.048 \\
J0410 & -0.006 & -0.008 & 0.128 & -0.152 & 0.097 \\
J0420 & 0.0 & -0.019 & 0.198 & -0.16 & 0.218 \\
J0430 & -0.028 & -0.032 & 0.062 & -0.078 & 0.035 \\
J0440 & -0.021 & -0.024 & 0.055 & -0.07 & 0.026 \\
J0450 & -0.011 & -0.023 & 0.113 & -0.114 & 0.121 \\
J0460 & -0.023 & -0.023 & 0.041 & -0.062 & 0.01 \\
J0470 & -0.024 & -0.028 & 0.045 & -0.076 & 0.019 \\
J0480 & -0.033 & -0.029 & 0.053 & -0.08 & 0.015 \\
J0490 & -0.015 & -0.013 & 0.071 & -0.098 & 0.051 \\
J0500 & -0.022 & -0.022 & 0.038 & -0.061 & 0.021 \\
J0510 & -0.023 & -0.018 & 0.035 & -0.06 & 0.015 \\
J0520 & -0.022 & -0.019 & 0.07 & -0.088 & 0.031 \\
J0530 & -0.022 & -0.018 & 0.036 & -0.068 & 0.014 \\
J0540 & -0.025 & -0.028 & 0.041 & -0.07 & 0.014 \\
J0550 & -0.028 & -0.031 & 0.036 & -0.074 & 0.02 \\
J0560 & -0.034 & -0.028 & 0.067 & -0.1 & 0.033 \\
J0570 & -0.031 & -0.029 & 0.052 & -0.073 & 0.02 \\
J0580 & -0.024 & -0.021 & 0.043 & -0.076 & 0.011 \\
J0590 & -0.021 & -0.024 & 0.051 & -0.067 & 0.029 \\
J0600 & -0.022 & -0.018 & 0.046 & -0.066 & 0.018 \\
J0610 & -0.015 & -0.018 & 0.053 & -0.066 & 0.03 \\
J0620 & -0.016 & -0.019 & 0.047 & -0.065 & 0.029 \\
J0630 & -0.016 & -0.018 & 0.055 & -0.083 & 0.036 \\
J0640 & -0.016 & -0.017 & 0.046 & -0.058 & 0.026 \\
J0650 & -0.023 & -0.02 & 0.04 & -0.064 & 0.011 \\
J0660 & -0.021 & -0.017 & 0.036 & -0.064 & 0.012 \\
J0670 & -0.026 & -0.017 & 0.041 & -0.064 & 0.009 \\
J0680 & -0.017 & -0.013 & 0.034 & -0.059 & 0.014 \\
J0690 & -0.021 & -0.017 & 0.051 & -0.067 & 0.032 \\
J0700 & -0.025 & -0.026 & 0.056 & -0.07 & 0.034 \\
J0710 & -0.018 & -0.013 & 0.031 & -0.063 & 0.013 \\
J0720 & -0.021 & -0.016 & 0.043 & -0.056 & 0.026 \\
J0730 & -0.017 & -0.016 & 0.079 & -0.064 & 0.034 \\
J0740 & -0.021 & -0.012 & 0.049 & -0.056 & 0.02 \\
J0750 & -0.017 & -0.016 & 0.055 & -0.062 & 0.029 \\
J0760 & -0.019 & -0.016 & 0.064 & -0.051 & 0.008 \\
J0770 & -0.019 & -0.019 & 0.06 & -0.079 & 0.037 \\
J0780 & -0.012 & -0.01 & 0.068 & -0.077 & 0.045 \\
J0790 & -0.026 & -0.025 & 0.073 & -0.076 & 0.019 \\
J0800 & -0.028 & -0.025 & 0.061 & -0.101 & 0.022 \\
J0810 & -0.022 & -0.014 & 0.078 & -0.075 & 0.043 \\
J0820 & -0.02 & -0.022 & 0.078 & -0.072 & 0.044 \\
J0830 & -0.012 & -0.02 & 0.118 & -0.094 & 0.083 \\
J0840 & -0.02 & -0.012 & 0.086 & -0.079 & 0.049 \\
J0850 & -0.028 & -0.024 & 0.093 & -0.072 & 0.056 \\
J0860 & -0.017 & -0.015 & 0.09 & -0.065 & 0.065 \\
J0870 & -0.019 & -0.018 & 0.077 & -0.075 & 0.047 \\
J0880 & -0.015 & -0.019 & 0.068 & -0.073 & 0.029 \\
J0890 & -0.032 & -0.027 & 0.072 & -0.112 & 0.026 \\
J0900 & -0.016 & -0.014 & 0.096 & -0.077 & 0.056 \\
J0910 & -0.036 & -0.022 & 0.113 & -0.105 & 0.043 \\
J1007 & -0.022 & -0.022 & 0.098 & -0.095 & 0.036 \\
uJPAS & -0.127 & -0.051 & 0.25 & -0.233 & 0.024 \\
gSDSS & -0.02 & -0.019 & 0.035 & -0.061 & 0.008 \\
rSDSS & -0.02 & -0.018 & 0.029 & -0.054 & 0.01 \\
iSDSS & -0.021 & -0.016 & 0.05 & -0.055 & 0.008 \\
\hline                  
\end{tabular}
\end{table*}

\begin{table*}
\caption{Statistical parameters of the difference $\Delta m_{AB}$ for the different filters, using the \magpetro \ photometry. The difference is calculated $\Delta \mathrm{MAG} = \mathrm{MAG}_{\sext} -\mathrm{MAG}_{\PyDJ} $. The parameters are, from left to right: mean, median, standard deviation, $10^\mathrm{th}$ percentile, and  $90^\mathrm{th}$ percentile.} 
\label{table:PETRO_wl}      
\centering          
\begin{tabular}{c c c c c c  }     
\hline\hline       
                      
Filter & mean $\Delta m_{AB}$ & median $\Delta m_{AB}$ &  $\sigma(\Delta m_{AB})$ & $P_{10}$ $\Delta m_{AB}$ &  $P_{90}$ $\Delta m_{AB}$\\ 
\hline                    
uJAVA & -0.079 & -0.098 & 0.389 & -0.462 & 0.181 \\
J0378 & -0.197 & -0.142 & 0.329 & -0.502 & 0.035 \\
J0390 & -0.073 & -0.084 & 0.244 & -0.298 & 0.163 \\
J0400 & -0.081 & -0.076 & 0.287 & -0.342 & 0.108 \\
J0410 & 0.012 & -0.021 & 0.309 & -0.291 & 0.333 \\
J0420 & 0.014 & -0.053 & 0.432 & -0.335 & 0.628 \\
J0430 & -0.055 & -0.077 & 0.224 & -0.333 & 0.159 \\
J0440 & -0.037 & -0.046 & 0.204 & -0.238 & 0.166 \\
J0450 & 0.003 & -0.019 & 0.285 & -0.297 & 0.429 \\
J0460 & -0.046 & -0.039 & 0.198 & -0.264 & 0.111 \\
J0470 & -0.039 & -0.058 & 0.213 & -0.283 & 0.107 \\
J0480 & -0.06 & -0.058 & 0.19 & -0.306 & 0.102 \\
J0490 & -0.022 & -0.048 & 0.234 & -0.321 & 0.162 \\
J0500 & -0.048 & -0.047 & 0.184 & -0.252 & 0.114 \\
J0510 & -0.042 & -0.034 & 0.177 & -0.286 & 0.06 \\
J0520 & -0.049 & -0.067 & 0.232 & -0.264 & 0.137 \\
J0530 & -0.046 & -0.029 & 0.188 & -0.303 & 0.049 \\
J0540 & -0.058 & -0.046 & 0.198 & -0.341 & 0.099 \\
J0550 & -0.06 & -0.05 & 0.194 & -0.269 & 0.061 \\
J0560 & -0.057 & -0.079 & 0.284 & -0.288 & 0.11 \\
J0570 & -0.072 & -0.053 & 0.215 & -0.308 & 0.072 \\
J0580 & -0.051 & -0.043 & 0.206 & -0.229 & 0.1 \\
J0590 & -0.064 & -0.04 & 0.221 & -0.296 & 0.104 \\
J0600 & -0.056 & -0.042 & 0.246 & -0.299 & 0.141 \\
J0610 & -0.053 & -0.036 & 0.224 & -0.281 & 0.187 \\
J0620 & -0.05 & -0.048 & 0.212 & -0.236 & 0.118 \\
J0630 & -0.042 & -0.038 & 0.219 & -0.316 & 0.146 \\
J0640 & -0.046 & -0.022 & 0.246 & -0.256 & 0.086 \\
J0650 & -0.06 & -0.024 & 0.234 & -0.275 & 0.079 \\
J0660 & -0.065 & -0.031 & 0.267 & -0.322 & 0.075 \\
J0670 & -0.077 & -0.028 & 0.27 & -0.361 & 0.075 \\
J0680 & -0.054 & -0.015 & 0.216 & -0.284 & 0.087 \\
J0690 & -0.058 & -0.034 & 0.248 & -0.238 & 0.087 \\
J0700 & -0.069 & -0.05 & 0.282 & -0.391 & 0.165 \\
J0710 & -0.06 & -0.04 & 0.23 & -0.243 & 0.116 \\
J0720 & -0.064 & -0.028 & 0.244 & -0.279 & 0.081 \\
J0730 & -0.055 & -0.052 & 0.37 & -0.213 & 0.226 \\
J0740 & -0.068 & -0.031 & 0.301 & -0.285 & 0.082 \\
J0750 & -0.042 & -0.022 & 0.277 & -0.339 & 0.149 \\
J0760 & -0.081 & -0.039 & 0.388 & -0.375 & 0.089 \\
J0770 & -0.114 & -0.049 & 0.519 & -0.282 & 0.121 \\
J0780 & -0.07 & -0.034 & 0.518 & -0.218 & 0.162 \\
J0790 & -0.095 & -0.052 & 0.437 & -0.35 & 0.124 \\
J0800 & -0.105 & -0.074 & 0.364 & -0.43 & 0.089 \\
J0810 & -0.09 & -0.072 & 0.388 & -0.404 & 0.13 \\
J0820 & -0.074 & -0.044 & 0.415 & -0.294 & 0.12 \\
J0830 & -0.078 & -0.042 & 0.482 & -0.296 & 0.246 \\
J0840 & -0.067 & -0.058 & 0.36 & -0.254 & 0.171 \\
J0850 & -0.059 & -0.032 & 0.326 & -0.179 & 0.101 \\
J0860 & -0.027 & -0.023 & 0.304 & -0.187 & 0.183 \\
J0870 & -0.035 & -0.026 & 0.25 & -0.173 & 0.159 \\
J0880 & -0.018 & -0.058 & 0.272 & -0.222 & 0.241 \\
J0890 & -0.077 & -0.05 & 0.3 & -0.376 & 0.149 \\
J0900 & -0.062 & -0.015 & 0.437 & -0.355 & 0.197 \\
J0910 & -0.062 & -0.044 & 0.285 & -0.354 & 0.205 \\
J1007 & -0.038 & -0.046 & 0.285 & -0.399 & 0.119 \\
uJPAS & -0.228 & -0.124 & 0.401 & -0.487 & 0.052 \\
gSDSS & -0.039 & -0.023 & 0.181 & -0.264 & 0.053 \\
rSDSS & -0.052 & -0.022 & 0.207 & -0.31 & 0.082 \\
iSDSS & -0.094 & -0.024 & 0.437 & -0.273 & 0.099 \\
\hline                  
\end{tabular}
\end{table*}

\begin{table*}
\caption{Statistical parameters of the difference $\Delta m_{AB}$ for different brightness bins, using the \magpetro \ photometry. The difference is calculated $\Delta \mathrm{MAG} = \mathrm{MAG}_{\sext} -\mathrm{MAG}_{\PyDJ} $. The parameters are, from left to right: mean, median, standard deviation, $10^\mathrm{th}$ percentile, and  $90^\mathrm{th}$ percentile.}             
\label{table:PETRO_bright}      
\centering          
\begin{tabular}{c c c c c c  }     
\hline\hline       
                       
Brightness bin & mean $\Delta m_{AB}$ & median $\Delta m_{AB}$ &  $\sigma(\Delta m_{AB})$ & $P_{10}$ $\Delta m_{AB}$ &  $P_{90}$ $\Delta m_{AB}$\\ 
\hline                    
$13 \le m_{AB} < 14$ & -0.046 & -0.046 & 0.026 & -0.07 & -0.015 \\
$14 \le m_{AB} < 15$ & -0.079 & -0.087 & 0.033 & -0.109 & -0.035 \\
$15 \le m_{AB} < 16$ & -0.031 & -0.079 & 0.201 & -0.126 & 0.071 \\
$16 \le m_{AB} < 17$ & -0.022 & -0.033 & 0.207 & -0.198 & 0.112 \\
$17 \le m_{AB} < 18$ & -0.035 & -0.014 & 0.261 & -0.284 & 0.13 \\
$18 \le m_{AB} < 19$ & -0.139 & -0.043 & 0.409 & -0.519 & 0.177 \\
$19 \le m_{AB} < 20$ & -0.12 & -0.078 & 0.408 & -0.478 & 0.264 \\
$20 \le m_{AB} < 21$ & 0.08 & -0.013 & 0.511 & -0.599 & 0.625 \\
$21 \le m_{AB} < 22$ & 1.53 & 1.53 & 0.0 & 1.53 & 1.53 \\
$22 \le m_{AB} < 23$ & -0.655 & -0.655 & 0.0 & -0.655 & -0.655 \\
\hline                  
\end{tabular}
\end{table*}

\begin{table*}
\caption{Statistical parameters of the difference $\Delta m_{AB}$ for different photometry. The difference is calculated $\Delta \mathrm{MAG} = \mathrm{MAG}_{\sext} -\mathrm{MAG}_{\PyDJ} $. The parameters are, from left to right: mean, median, standard deviation, $10^\mathrm{th}$ percentile, and  $90^\mathrm{th}$ percentile.}             
\label{table:delta_phot}      
\centering          
\begin{tabular}{c c c c c c  }      
\hline\hline       
                      
Photometry & mean $\Delta m_{AB}$ & median $\Delta m_{AB}$ &  $\sigma(\Delta m_{AB})$ & $P_{10}$ $\Delta m_{AB}$ &  $P_{90}$ $\Delta m_{AB}$\\ 
\hline                    
\magauto & -0.021 & -0.01 & 0.082 & -0.075 & 0.027 \\
\magpetro &  -0.05 & -0.005 & 0.274 & -0.273 & 0.113 \\
MAG\_APER\_0\_8 & 0.005 & 0.0 & 0.061 & -0.023 & 0.049 \\
MAG\_APER\_1\_0 & 0.005 & 0.0 & 0.045 & -0.017 & 0.042 \\
MAG\_APER\_1\_2 & 0.005 & 0.001 & 0.038 & -0.016 & 0.035 \\
MAG\_APER\_1\_5 & 0.003 & 0.0 & 0.032 & -0.014 & 0.028 \\
MAG\_APER\_2\_0 & 0.0 & 0.0 & 0.027 & -0.015 & 0.021 \\
MAG\_APER\_3\_0 & -0.004 & 0.0 & 0.026 & -0.019 & 0.011 \\
MAG\_APER\_4\_0 & -0.006 & -0.0 & 0.029 & -0.022 & 0.008 \\
MAG\_APER\_6\_0 & -0.011 & -0.001 & 0.042 & -0.032 & 0.006 \\
\hline                  
\end{tabular}
\end{table*}

\end{document}